\documentclass[amssymb,amsmath,twocolumn,showpacs,prb]{revtex4}

\usepackage{color,graphicx,bm}
\usepackage{overpic}
\usepackage{textcomp}

\newcommand{\ua}{\uparrow}
\newcommand{\da}{\downarrow}
\def\k{{\bf k}}
\def\q{{\bf q}}
\newcommand{\mf}{^{\scalebox{0.5}{M$\!$F}}}

\begin{document}

\title{Superconductivity with Rashba Spin-Orbit Coupling and Magnetic Field\\[0mm]
}

\author{
Florian Loder$^{\,1,2}$, Arno P. Kampf$^{\,2}$, and Thilo Kopp$^{\,1}$
\vspace{0,3cm}}

\affiliation{Center for Electronic Correlations and Magnetism, $^1$Experimental Physics VI, $^2$Theoretical Physics III\\ 
Institute of Physics, University of Augsburg, 86135 Augsburg, Germany}

\date{\today}

\begin{abstract}
Two-dimensional electron systems at oxide interfaces are often influenced by a Rashba type spin-orbit coupling, which is tunable by a transverse electric field. Ferromagnetism near the interface can simultaneously induce strong local magnetic fields. This combination of spin-orbit coupling and magnetism leads to asymmetric two-sheeted Fermi surfaces, on which either intra- or inter-band pairing is favored. The superconducting order parameters are derived within a microscopic pairing model realizing both, the Bardeen-Cooper-Schrieffer superconductor with inter-band pairing, and a mixed-parity state with finite-momentum intra-band pairing. We present a phase diagram for the superconducting groundstates and analyze the density of states, the spectra, and the momentum distribution functions of the different phases. The results are discussed in the context of superconductivity and ferromagnetism at LaAlO$_3$-SrTiO$_3$ interfaces and superconductors with broken inversion symmetry.
\end{abstract}

\pacs{74.78.-w,74.25.N-,74.20.Rp}

\maketitle

\section{Introduction}
Reducing the symmetry of a superconducting system changes the properties of the possible superconducting (SC) states. In particular, if the inversion symmetry is broken, the combination of Rashba spin-orbit coupling (SOC) and magnetic fields leads to new classes of SC states. Superconductivity with SOC was first discussed in inversion-symmetry breaking heavy fermion systems~\cite{sigrist:91,pfleiderer:09}, and this topic recently expanded in the context of topological superconductors~\cite{yada11} and in normal--SC heterostructures~\cite{sau10,alicea10}.
With the discovery of superconductivity at the interface between LaAlO$_3$ (LAO) and SrTiO$_3$ (STO)~\cite{reyren07,caviglia08}, a system was identified which manifestly exhibits a combination of SOC, magnetism, and superconductivity.
At this interface, a dilute, almost ideally two-dimensional electron liquid forms~\cite{ohtomo04,thiel06,sing09}. The polarity of the interface induces a perpendicular electric field $\bf E$ and gives rise to a Rashba SOC $\gamma({\bf E}\times\hat{\bf p})\cdot\hat{\bf S}$, where $\hat{\bf p}$ and $\hat{\bf S}$ are momentum and spin operators, respectively. Quantitative estimates for the SOC parameter $\gamma$ are under debate~\cite{caviglia10,shalom:10,fete:12,joshua:11,zhong:13}. Recent experiments also revealed inhomogeneous in-plane magnetism at the interface~\cite{dikin11,li11,bert11} which coexists with superconductivity at low temperatures and is possibly generated through oxygen vacancies~\cite{pavlenko11,pavlenko12}. A physically similar situation may also be realized in thin films of heavy-fermion superconductors in an external magnetic field. It is this special concurrence of superconductivity with magnetism and SOC which allows for the formation of unanticipated multi-component SC states. 

The magnetic field ${\bf B}$ couples through a Zeeman term $\mu_{\rm B}{\bf B}\cdot\hat{\bf S}$ to the spin in the same way as ${\bf E}\times\hat{\bf p}$ through the Rashba SOC. Both couplings result in a band splitting and a two-sheeted Fermi surface. Various studies focused on the emergence of topological edge states in two-dimensional systems with ${\bf B}\parallel{\bf E}$~\cite{sato09,sato10}. However, interesting new physics emerges in a boundary free system with an in-plane field ${\bf B}\perp{\bf E}$. An in-plane field does not generate vortices in the superconducting state, but couples to the electrons through a Zeeman term.
While for ${\bf B}=\bm0$ intra-band pairing with zero center-of-mass momentum (COMM) is expected, a finite COMM is required if the in-plane component of the magnetic-field exceeds a certain limit~\cite{barzykin02,michaeli12}. If the magnetic field dominates over SOC, a crossover to inter-band pairing must occur, of either BCS type with zero COMM for low fields, or a finite COMM state above a critical field, as proposed by Fulde and Ferrell~\cite{fulde64} and by Larkin and Ovchinnikov~\cite{larkin64}. The regime with both SOC and magnetic field has previously been treated within a Ginzburg-Landau analysis for the ``helical phase'' of the intra-band pairing regime~\cite{mineev93}, which was later extended to other non inversion-symmetric systems~\cite{kaur05,dimitrova07}. Surface superconductivity in magnetic fields was discussed also on the basis of the linearized Gor'kov equations~\cite{barzykin02, tanaka07}.

In this paper we characterize the possible SC groundstates in the presence of both SOC and in-plane magnetic field within a microscopic model. We describe the transition from the intra-band pairing regime with dominant SOC to the inter-band regime with a dominant Zeeman coupling. These two regimes are typically separated by a first-order phase boundary and exhibit distinctly different signatures in the density of states (DOS). On the basis of existing experimental data for the LAO-STO interface we discuss the implications for the SC state which may be obtained from tunneling spectra.

\section{The Normal State}
For the microscopic description of a two-dimensional electron system we use a single-band tight-binding model on a square lattice with $N$ sites. We include a nearest-neighbor hopping $t$ and a next-nearest-neighbor hopping $t'$ with a kinetic energy of the form ${\cal H}_0=\sum_{\k,s}\epsilon_\k c^\dag_{\k s}c_{\k s}$ with $\epsilon_\k=-2t(\cos k_x+\cos k_y)+4t'\cos k_x\cos k_y-\mu$; $\mu$ is the chemical potential.

\begin{figure}[t!]
\centering
\vspace{3mm}
\begin{overpic}
[width=0.99\columnwidth]{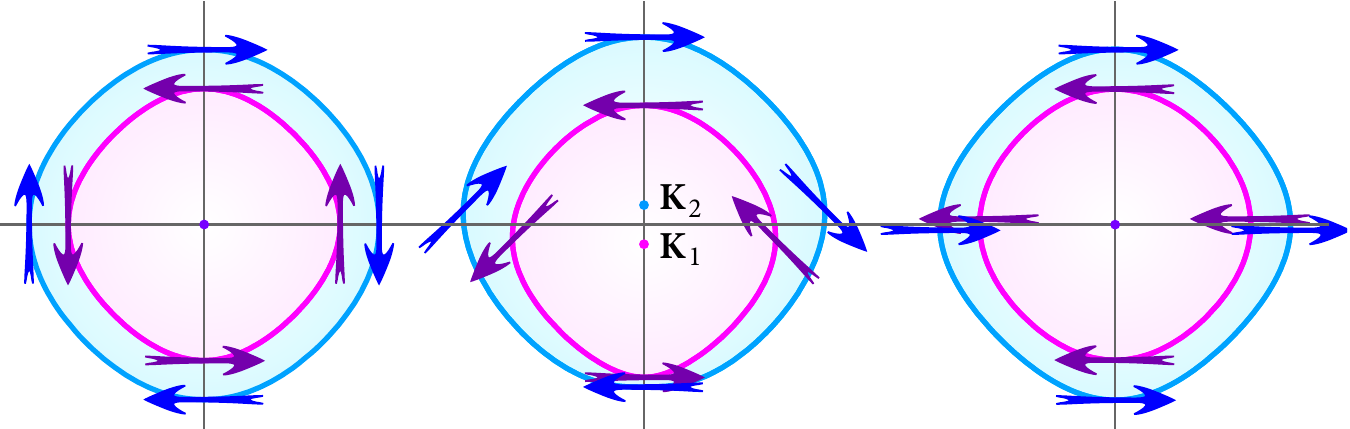}
\put(-2,29){(a)}
\put(31,29){(b)}
\put(67,29){(c)}
\end{overpic}
\vspace{-2mm}
\caption{(Color online) Fermi surfaces with SOC and magnetic field for the tight-binding dispersion $\epsilon_\k$ and band filling $n=0.24$. (a) $\alpha=0.4\,t$, $H_x=0$. (b) $\alpha=0.4\,t$, $H_x=0.3\,t$. (c) $\alpha=0$, $H_x=0.4\,t$.
}
\label{Fig1}
\end{figure}

The spin part ${\cal H}_{\rm S}$ of the Hamiltonian consists of the Rashba SOC and the Zeeman coupling to an in-plane magnetic field.
With the spin operator $\hat{\bf S}_i=\sum_{s,s'}c^\dag_{is}\bm\sigma_{ss'}c_{is'}/2$ on the lattice site $i$, ${\cal H}_{\rm S}$ becomes in momentum space
\begin{align}
{\cal H}_{\rm S}=\sum_{\k,s}(A_{\k,s}+H_x+isH_y)\,c^\dag_{\k,s}c_{\k,-s}
\label{g1}
\end{align}
with $A_{\k,s}=\alpha(\sin k_y +is\sin k_x)$, $\alpha=\hbar^2|{\bf E}|\gamma/2$, ${\bf H}=\mu_{\rm B}\bf B$, and $s=\pm1$. The effective electron mass is denoted by $m^{\!*}$. 
For free electrons moving in a perpendicular electric field, the parameter $\gamma=e/(2{m^*}^2c^2)$ derives directly from the Dirac Hamiltonian. However, the effective value of $\gamma$ can be strongly enhanced in multi-band systems with atomic SOC~\cite{joshua:11,zhong:13}.

The term ${\cal H}_\text S$ generates superpositions of electronic states with spin $\ua$ and $\da$ and equal momentum $\k$, represented by the fermionic operators
$a_{\k1}=c_{\k\ua}+\phi_\k c_{\k\da}$ and $a_{\k2}=c_{\k\ua}-\phi_\k c_{\k\da}$
with the phase factor
\begin{align}
\phi_\k=\frac{A_{\k\ua}+H_x+iH_y}{|A_{\k\ua}+H_x+iH_y|}.
\label{g4}
\end{align}
A momentum dependent in-plane spin orientation results.
The indices 1 and 2 denote the two energy bands $\xi_{\k1}$ and $\xi_{\k2}$ resulting from the non-interaction Hamiltonian, which is diagonal in the operators $a_{\k1}$ and $a_{\k2}$:
\begin{align}
{\cal H}_0+{\cal H}_{\rm S}=\sum_\k\left(\xi_{\k1}a^\dag_{\k1}a_{\k1}+\xi_{\k2}a^\dag_{\k2}a_{\k2}\right)
\label{g5}
\end{align}
with $\xi_{\k1,2}=\epsilon_\k\pm|A_{\k\ua}+H_x+iH_y|$.

The two corresponding Fermi surfaces are shown in Fig.~\ref{Fig1} for a charge density $n=0.24$. Three qualitatively different cases are identified: (a) For finite SOC ($\alpha\neq0$) but no magnetic field ($H=0$), two concentric Fermi surfaces emerge, on which the spins are oriented anti-parallel ($\xi_{\k1}$) or parallel ($\xi_{\k2}$) to ${\bf E}\times\bm\nabla\epsilon_\k$, i.e. the spin winds once around its quantization axis upon circulating the Fermi surface. (c) In a magnetic field (here $\bf H$ in $x$-direction) with vanishing SOC ($\alpha=0$), the Fermi surfaces are also concentric, however with spins oriented anti-parallel ($\xi_{\k1}$) or parallel ($\xi_{\k2}$) to $\bf H$. In the following we shall mostly analyze the intermediate regime with both $\alpha\neq0$ and $H_x\neq0$ shown in (b). The Fermi surfaces are displaced from the Brillouin-zone center perpendicular to $\bf H$, e.g. to the positions ${\bf K}_1=(0,K_1)$ and ${\bf K}_2=(0,K_2)$ for $H_y=0$. The offsets are maximal on the critical line $\alpha=H_x$, and in general $|K_1| > |K_2|$. For small magnetic fields $H_x\ll\alpha$, the offsets are given by $K_1=K_2\approx H_x/\sqrt{\alpha^2+4t|\mu|}$ in first order of $H_x$ (c.f.~Refs.~[\onlinecite{barzykin02,michaeli12}]). The spin-winding number switches from zero for $\alpha<H_x$ to $\pm1$ for $\alpha>H_x$. A first-order transition between two superconducting states of different symmetry class is therefore expected at this crossover.

\section{The Pairing Interaction}

In case (a), intra-band pairing of anti-parallel spins with zero COMM is expected on both Fermi surfaces. Although spin-triplet inter-band pairing with a finite COMM is allowed in the presence of SOC, it is unlikely to be energetically favorable. In (c) only singlet pairing is possible for a local attraction. This requires inter-band pairing and allows for the formation of a Larkin-Ovchinnikov (LO) state with finite COMMs $\q$ and $-\q$, where $\q$ connects the two Fermi surfaces, if $|{\bf H}|$ is larger than the SC order parameter~\footnote{A Fulde-Ferrell state with only one $\q$, is also possible, but energetically less favorable.}.
Which pairing type is realized in the intermediate regime (b) with $\alpha\neq0$ and $H_x\neq0$ is difficult to anticipate. Nevertheless, if $H_x\ll\alpha$ we expect an intra-band pairing state where finite COMMs $\q_1$ and $\q_2$ are realized which optimize pairing on band 1 and band 2, respectively. These COMMs should be chosen such that the momenta of the paired eigenstates lie on the Fermi surface of the corresponding band, i.e. $\q_1\approx2{\bf K}_1$ and $\q_2\approx2{\bf K}_2$. (The exact values of $\q_1$ and $\q_2$ can only be found from a self-consistent numerical calculation and may deviate slightly from the expectations formulated above.) Such a state was considered in Ref.~[\onlinecite{kaur05}], where it was assumed that only one COMM exists which optimizes pairing on the larger Fermi surface.

In the following we analyze the pairing originating from an on-site pairing interaction and characterize the crossover from intra- to inter-band pairing for increasing field $H_x$. The interaction term in momentum space has the form
\begin{align}
{\cal H}_{\rm I}=-\frac{V}{2N^2}\sum_{\k,\k'}\sum_\q\sum_sc^\dag_{\k,s}c^\dag_{-\k+\q,-s}c_{-\k'+\q,-s}c_{\k',s}.
\label{g2}
\end{align}
Expressing ${\cal H}_{\rm I}$ in terms of the band operators $a_{\k,1}$ and $a_{\k,2}$ leads to
\begin{multline}
{\cal H}_{\rm I}=\frac{V}{2N^2}\sum_{\k,\k',\q}[V_{\text{even}}(\k,\k',\q)+V_{\text{odd}}(\k,\k',\q)]\\
\times\sum_{\alpha,\beta,\alpha',\beta'}\gamma_{\beta\beta'}\Big[a^\dag_{\k\alpha}a^\dag_{-\k+\q\beta}a_{-\k'+\q\beta'}a_{\k'\!\alpha'}\Big],
\label{g2.1}
\end{multline}
where $\gamma_{\beta\beta'}=1$ for $\beta=\beta'$ and $\gamma_{\beta\beta'}=-1$ for $\beta\neq\beta'$.

In the operators $a_{\k1}$ and $a_{\k2}$, the interaction is non-local and acquires the momentum dependence of $\phi_\k$ (with $p$-wave symmetry for ${\bf H}=\bm0$). The interaction is decomposed into a component $V_{\text{even}}(\k,\k',\q)=Vg^*_{\text{even}}(\k,\q)g_{\text{even}}(\k',\q)$ which is even under the permutation $\k\leftrightarrow-\k+\q$ and a component $V_{\text{odd}}(\k,\k',\q)=Vg^*_{\text{odd}}(\k,\q)g_{\text{odd}}(\k',\q)$ which is odd under the same permutation, where
\begin{align}
g_{\text{even}}(\k,\q)&=\left(\phi_{\k}+\phi_{-\k+\q}\right)/2,
\label{g8}\\
g_{\text{odd}}(\k,\q)&=\left(\phi_{\k}-\phi_{-\k+\q}\right)/2.
\label{g9}
\end{align}
The mean-field pairing order parameters (OPs) $\Delta_{\alpha\beta}(\k,\q)$ generated by the interaction~(\ref{g2.1}) must be antisymmetric under an exchange of the paired quasiparticles, i.e.
\begin{align}
\Delta_{\alpha\beta}(\k,\q)=-\Delta_{\beta\alpha}(-\k+\q,\q).
\end{align}
From this condition follows that the two intra-band OPs are of odd parity:
\begin{align}
\Delta_{11}(\k,\q)&=\frac{V}{N}\sum_{\k'}V_{\text{odd}}(\k,\k',\q)\langle a_{-\k'+\q1}a_{\k'1}\rangle,
\label{g12}\\
\Delta_{22}(\k,\q)&=\frac{V}{N}\sum_{\k'}V_\text{odd}(\k,\k',\q)\langle a_{-\k'+\q2}a_{\k'2}\rangle,
\label{g13}
\end{align}
whereas inter-band pairing allows for odd and even parity OPs:
\begin{align}
\Delta_{12}^\text{odd}(\k,\q)&=\Delta_{21}^\text{odd}(\k,\q)\nonumber\\&=\frac{V}{N}\sum_{\k'}V_{\text{odd}}(\k,\k',\q)\langle a_{-\k'+\q1}a_{\k'2}\rangle,
\label{g10}\\
\Delta_{12}^\text{even}(\k,\q)&=-\Delta_{21}^\text{even}(\k,\q)\nonumber\\&=\frac{V}{N}\sum_{\k'}V_{\text{even}}(\k,\k',\q)\langle a_{-\k'+\q1}a_{\k'2}\rangle.
\label{g11}
\end{align}
In the mean-field decoupling of Eq.~(\ref{g2.1}), all OPs coupling to the intra-band term $a^\dag_{\k'\alpha}a^\dag_{-\k'+\q\alpha}$ must be of odd parity with the OP $\Delta_{12}^\text{even}(\k,\q)$ canceling out. The inter-band term $a^\dag_{\k'\alpha}a^\dag_{-\k'+\q\beta}$ ($\alpha\neq\beta$) allows for both, odd and even parity OPs. However, all odd parity OPs cancel out in these terms. The mean-field pairing interaction therefore takes the form
\begin{align}
\begin{split}
{\cal H}\mf_{\rm I}=\frac{1}{N}\sum_{\k,\q}\Big[&\Delta^\text{odd}(\k,\q)\left(a^\dag_{\k1}a^\dag_{-\k+\q1}-a^\dag_{\k2}a^\dag_{-\k+\q2}\right)\\
+&\Delta^\text{even}(\k,\q)\left(a^\dag_{\k1}a^\dag_{-\k+\q2}-a^\dag_{\k2}a^\dag_{-\k+\q1}\right)\Big]
\label{g7}
\end{split}
\end{align}
with
\begin{align}
\Delta^\text{odd}(\k,\q)&=[\Delta_{11}(\k,\q)-\Delta_{22}(\k,\q)]/2,\\
\Delta^\text{even}(\k,\q)&=\Delta_{12}^\text{even}(\k,\q).
\label{g7.1}
\end{align}
These OPs may be expressed through the $\k$-independent quantities
\begin{align}
\Delta^\text{even/odd}(\q)=g_\text{even/odd}(\k,\q)\Delta^\text{even/odd}(\k,\q),
\label{g7.1}
\end{align}
since $g_\text{even/odd}(\k,\q)$ cancels the $\k$-dependence in $\Delta^\text{even/odd}(\k,\q)\propto g_\text{even/odd}^*(\k,\q)$. In the limit $H_x\rightarrow0$ one finds $\Delta_{11}(\k,\q)=-\Delta_{22}(\k,\q)$ and $\Delta^\text{even}(\k,\q)=0$, thus a pure odd-parity OP is realized.

\begin{table*}[ht]
\begin{tabular}{|c|c|c||c|c||c|c|c|c|}
\hline intra-band&$\alpha$&$H_x$&$\q_1$&$\q_2$&$\Delta_{11}(\q_1)$&$\Delta_{22}(\q_1)$&$\Delta_{11}(\q_2)$&$\Delta_{22}(\q_2)$\\\hline\hline
a1&$0.4$&$0.1$&$(0,0)$&$(0,0)$&$-0.234$&$0.297$&$-0.234$&$0.297$\\\hline
a2&$0.4$&$0.223$&$(0,0)$&$(0,0)$&$-0.132$&$0.178$&$-0.132$&$0.178$\\\hline
b1&$0.4$&$0.25$&$(0,-21)$&$(0,18)$&$-0.060$&$0.044$&$-0.048$&$0.109$\\\hline
b2&$0.4$&$0.3$&$(0,-28)$&$(0,27)$&0.002&0.001&$-0.012$&$-0.035$\\\hline\hline
inter-band&$\alpha$&$H_x$&$\q_1$&$\q_2$&$\Delta_{12}^\text{even}(\q_1)$&$\Delta_{12}^\text{odd}(\q_1)$&$\Delta_{12}^\text{even}(\q_2)$&$\Delta_{12}^\text{odd}(\q_2)$\\\hline\hline
c&$0.1$&$0.2$&$(0,0)$&$(0,0)$&$0.225$&--&$0.225$&--\\\hline
d&$0.05$&$0.44$&$(-52,0)$&$(52,0)$&$0.080$&--&$0.080$&--\\\hline
\end{tabular}
\caption{Summary of solutions from the four distinct SC phases for $V=1.7\,t$ (except 2-band b: $V=2.4\,t$), $t'=0$, and $n=0.24$, calculated on a 640$\times$640 lattice. For intra-band pairing, $\Delta_{11}(\q_{1,2})$ and $\Delta_{22}(\q_{1,2})$ correspond to pair-momenta $\q_{1,2}$, whereas for inter-band pairing, as well as for the LO solution, only the OPs $\Delta_{12}^\text{even}(\q_{1,2})$ exist. $\alpha$, $H$ and all OPs are given in units of $t$ and $\q_{1,2}$ in units of $2\pi/640\,a$.}
\label{Tab1}
\end{table*}

The OPs $\{\Delta_{11}(\k,\q),\Delta_{22}(\k,\q),\Delta_{12}^\text{odd}(\k,\q)\}$ form a triplet in the band space spanned by $a_{\k1}$ and $a_{\k2}$, and $\Delta_{12}^\text{even}(\k)$ is a band-space singlet. They can be transformed back to the standard spin singlet $\psi(\k,\q)$ and triplet ${\bf d}(\k,\q)=(d_x(\k,\q),d_y(\k,\q),d_z(\k,\q))$ OPs~\cite{sigrist:91,yanase07} using the inverse transformation $a_{\k i}\rightarrow c_{\k s}$.
In this way one finds that
\begin{align}
\psi(\k,\q)&=g_\text{even}(\k,\q)\Delta^\text{even}(\k,\q)+g_\text{odd}(\k,\q)\Delta^\text{odd}(\k,\q)\nonumber\\
&=\Delta^\text{even}(\q)+\Delta^\text{odd}(\q)
\end{align}
and
\begin{align}
d_z(\k,\q)=g_\text{odd}(\k,\q)\Delta^\text{even}(\k,\q)+g_\text{even}(\k,\q)\Delta^\text{odd}(\k,\q),
\end{align}
whereas the equal-spin OPs $d_x(\k,\q)$ and $d_y(\k,\q)$ are absent for the local pairing interaction. 
Singlet pairing is thus realized if either $g_\text{even}(\k,\q)$ or $g_\text{odd}(\k,\q)$ is zero, i.e., if either $\alpha$ or $H_x$ is zero. In this case, the parity of the OP is either purely even or purely odd. If both, $\alpha$ and $H_x$ are finite, the parity is mixed and therefore a spin-triplet component is admixed.

A phase-coherent superconducting state is realized if one (or a few) COMMs are macroscopically occupied. We choose a test set of COMMs according to the expectations on pairing we formulated above, i.e., two COMMs $\q_1$ and $\q_2$ should be present to optimize intra-band pairing in band 1 and band 2, and two additional COMMs $\pm\q$ which are required to form an LO state. Further, the COMM $\q=\bm0$ is likely to represent the groundstate for small values of $\alpha$ and $H_x$ and is therefore included as well. Possibly existing higher harmonics of these vectors are neglected here.

\begin{figure}[h]
\centering
\vspace{3mm}
\begin{overpic}
[width=\columnwidth]{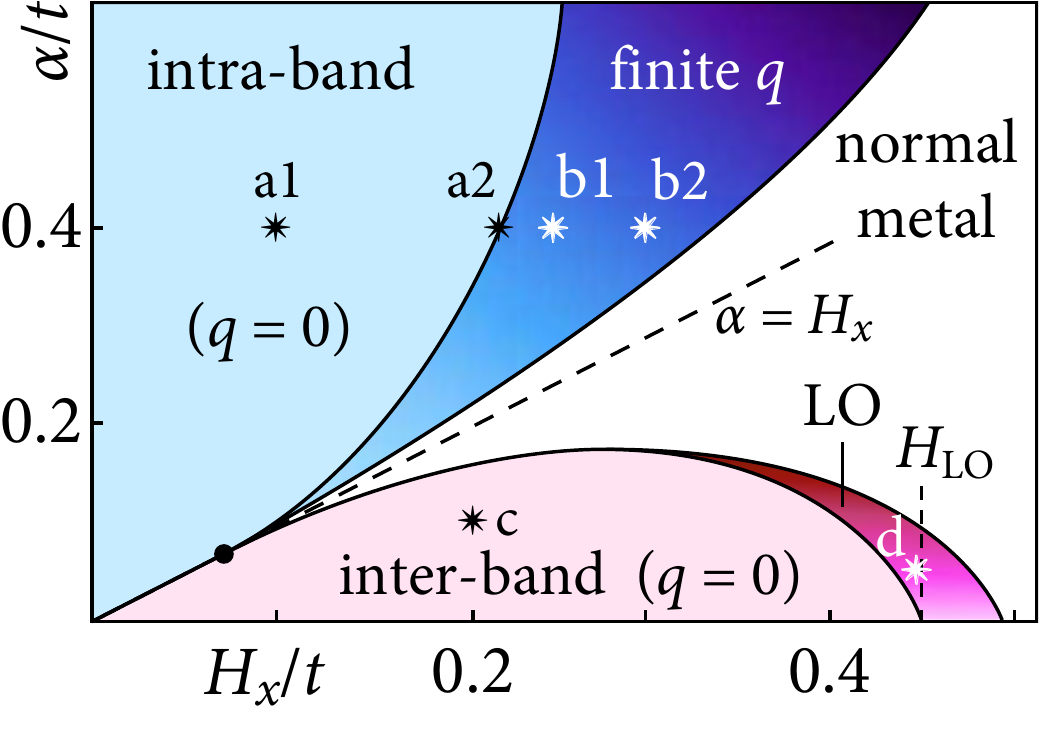}
\end{overpic}
\vspace{-3mm}
\caption{(Color online) Qualitative phase diagram for the model defined by the Hamiltonian~(\ref{g7}) for $V=1.7\,t$ and $t'=0.2\,t$, and $n=0.24\,t$. The blue regions in the upper-left half indicate intra-band pairing with dominating SOC, whereas the pink regions in the lower-right half represent inter-band pairing with either $q_x=0$ or $q_x\neq0$ (LO). Darker color in direction parallel to the $\alpha=H_x$ line indicates an increasing pair momentum $q_y$. The stars mark the positions of the five examples listed in Table~\ref{Tab1}. The critical magnetic field $H_\text{LO}=\Delta^\text{even}(H_x=0)$ marks the breakdown field of the classical BCS superconductor.
}
\label{Fig2}
\end{figure}

In the presence of $Q$ different COMMs $\q_i$ in the test set, the Hamiltonian ${\cal H}\mf={\cal H}_0+{\cal H}_{\rm S}+{\cal H}\mf_{\rm I}$ is expressed by a $2(Q+1)\times2(Q+1)$ matrix as
\begin{align}
\scalebox{0.92}[1]{$\displaystyle{
{\cal H}\mf=\sum_{\k}{\bf A}^\dag_\k\underbrace{\begin{pmatrix}\hat E_\k&\hat\Delta_{\k,\q_1}&\cdots&\cdots&\hat\Delta_{\k,\q_Q}&\cr\hat\Delta^\dag_{\k,\q_1}&-\hat E_{-\k+\q_1}&0&\cdots&0\cr\vdots&0&\ddots&0&\vdots\cr\vdots&\vdots&0&\ddots&0\cr\hat\Delta_{\k,\q_Q}&0&\cdots&0&-\hat E_{-\k+\q_Q}\end{pmatrix}}_{\displaystyle{\hat H_\k}}{\bf A}_\k}$}
\label{s6}
\end{align}
with
$
{\bf A}^\top_\k=(a_{\k1}, a_{\k2}, a^\dag_{-\k+\q_11}, a^\dag_{-\k+\q_12},\dots, a^\dag_{-\k+\q_Q1},$ $a^\dag_{-\k+\q_Q2}).
$
Here, $\hat E_\k$ and $\hat\Delta_{\k,\q}$ are the $2\times2$ matrices
\begin{align}
\hat E_\k&=\begin{pmatrix}\xi_{\k1}&0\cr0&\xi_{\k2}\end{pmatrix}
\label{s7.0}
\end{align}
and
\begin{align}
\scalebox{0.92}[1]{$\displaystyle{
\hat\Delta_{\k,\q}=\begin{pmatrix}\Delta^\text{odd}(\k,\q)&\Delta^\text{even}(\k,\q)\cr-\Delta^\text{even}(\k,\q)&-\Delta^\text{odd}(\k,\q)\end{pmatrix}.}$}
\label{s7}
\end{align}
${\cal H}\mf$ has to be diagonalized numerically to calculate the OPs $\Delta^\text{even/odd}(\q_i)$ self-consistently. Below we analyze these results and compare them to the above expectations in terms of an $\alpha$--$H_x$ phase diagram (Fig.~\ref{Fig2}).

\begin{figure*}[t]
\centering
\vspace{3mm}
\begin{overpic}
[width=0.95\columnwidth]{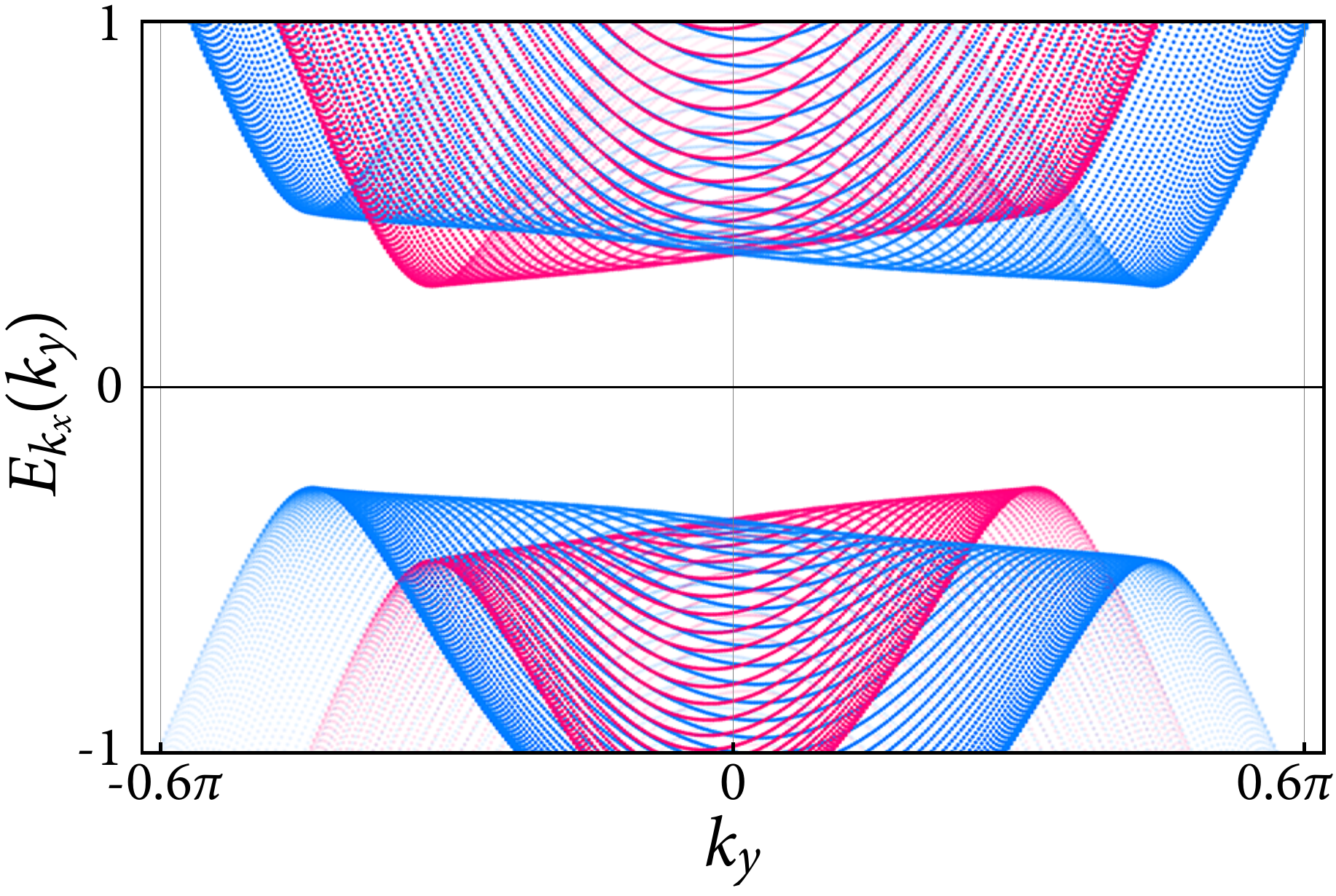}
\put(-5,63.5){(a1)}
\end{overpic}
\hspace{5mm}
\begin{overpic}
[width=0.95\columnwidth]{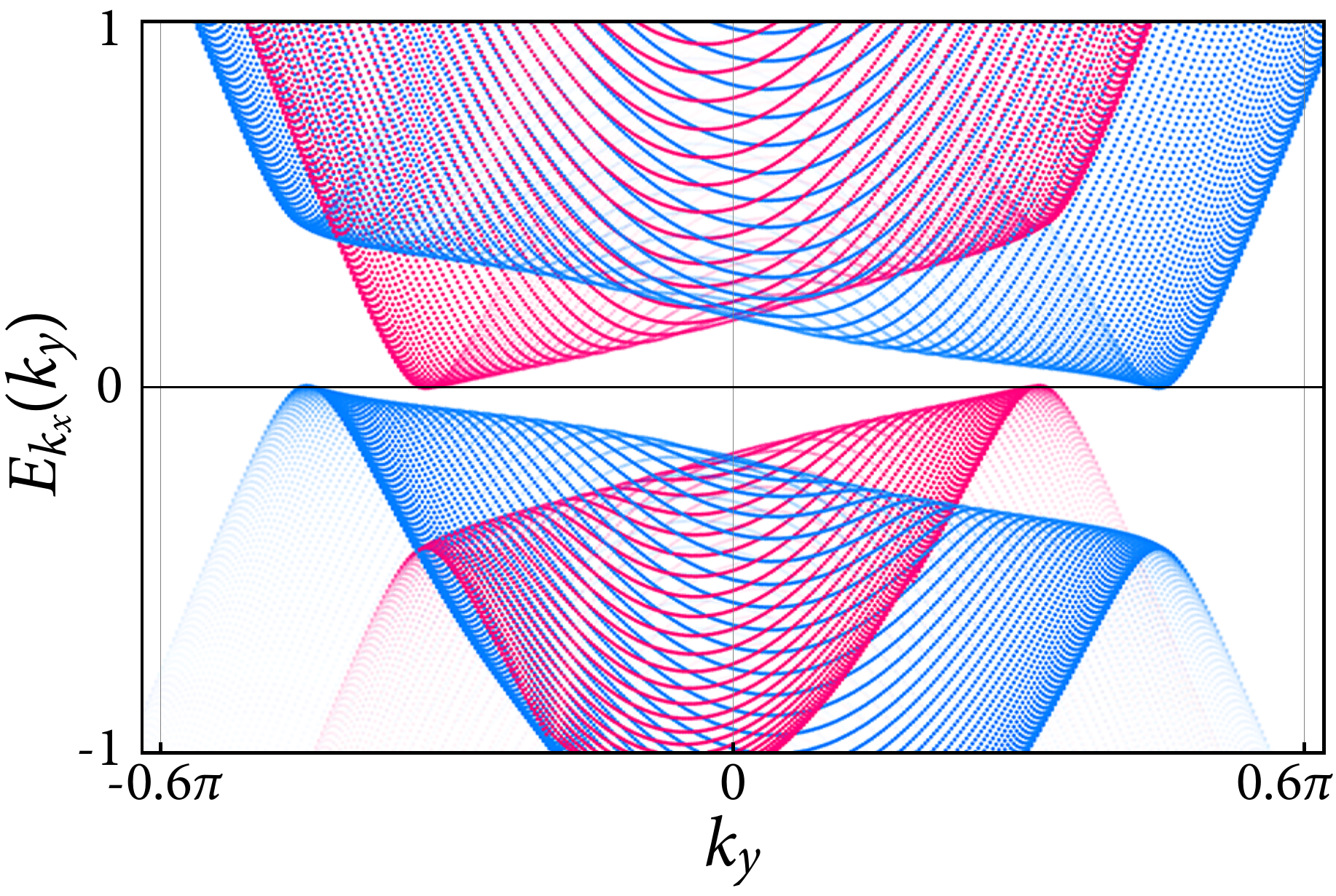}
\put(-5,63.5){(a2)}
\end{overpic}
\begin{overpic}
[width=0.95\columnwidth]{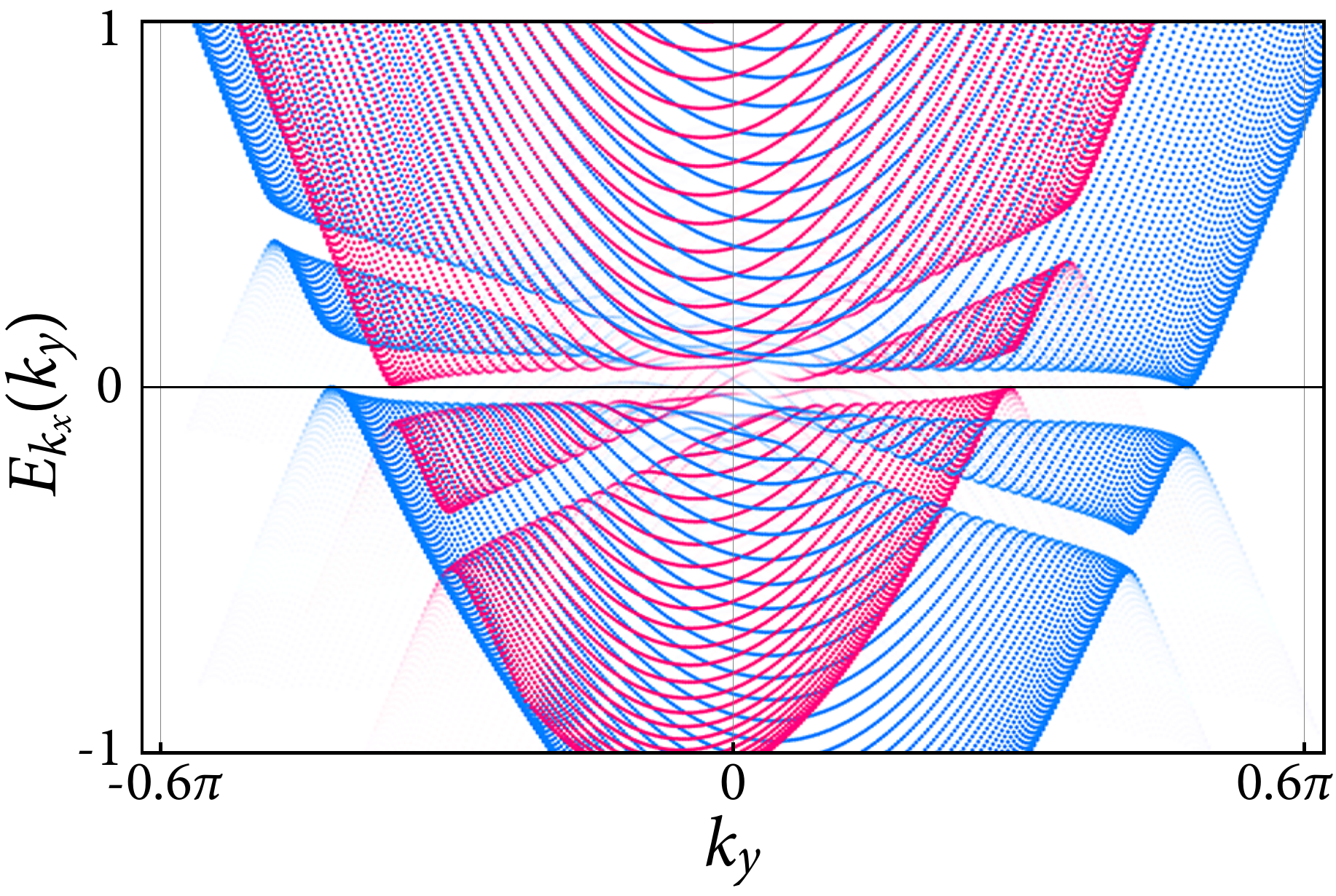}
\put(-5,63.5){(b1)}
\end{overpic}
\hspace{5mm}
\begin{overpic}
[width=0.95\columnwidth]{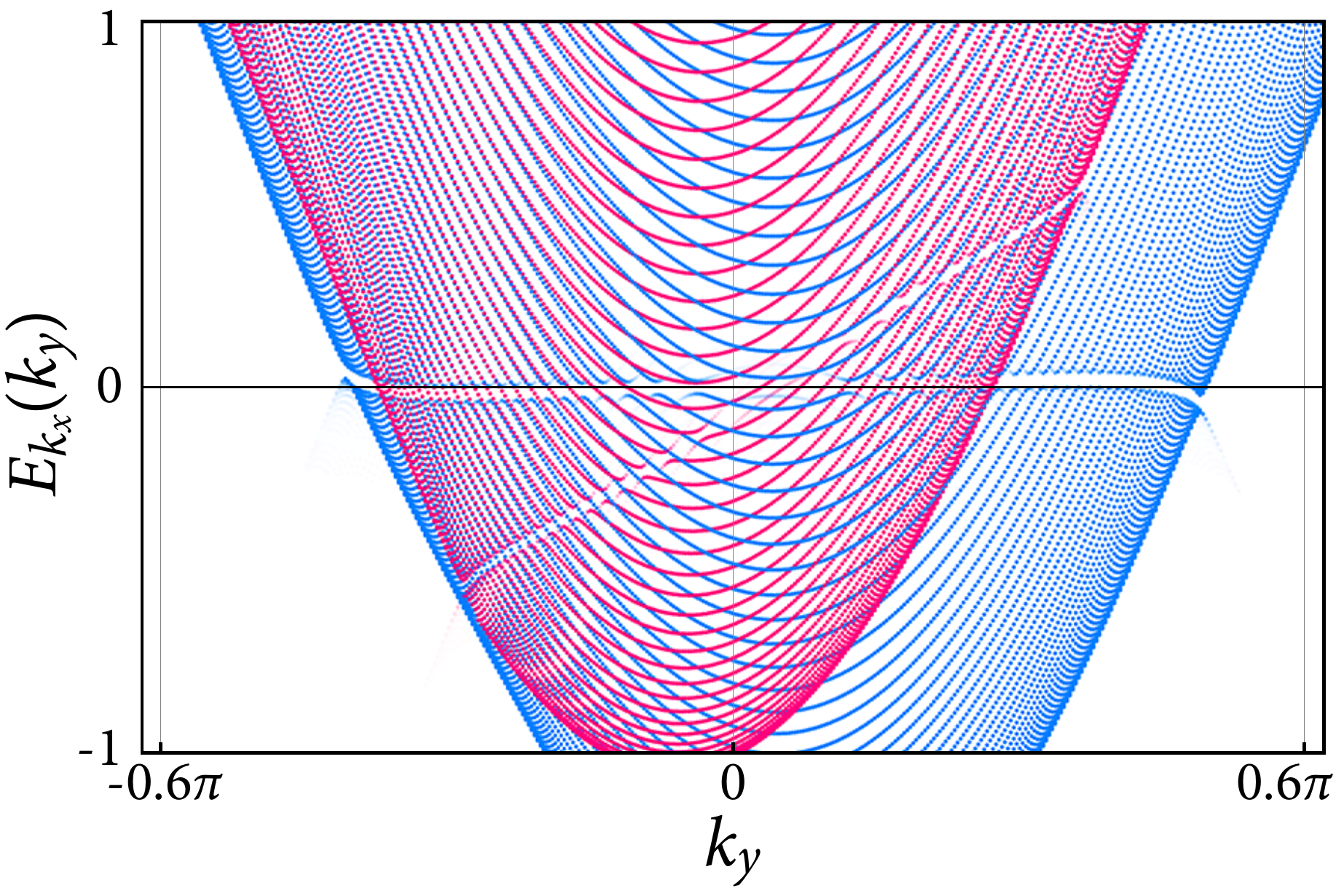}
\put(-5,63.5){(b2)}
\end{overpic}
\begin{overpic}
[width=0.95\columnwidth]{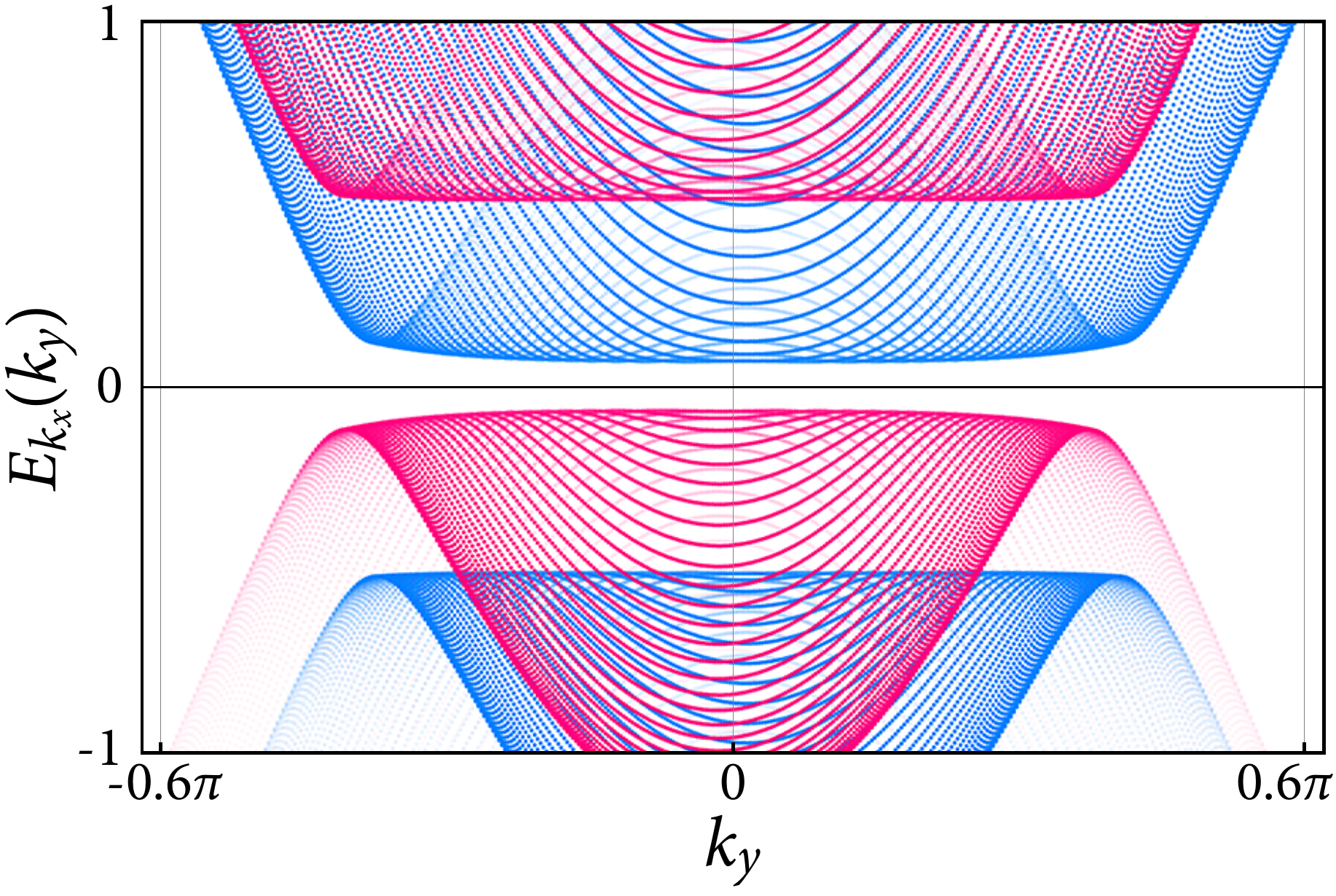}
\put(-5,63.5){(c)}
\end{overpic}
\hspace{5mm}
\begin{overpic}
[width=0.95\columnwidth]{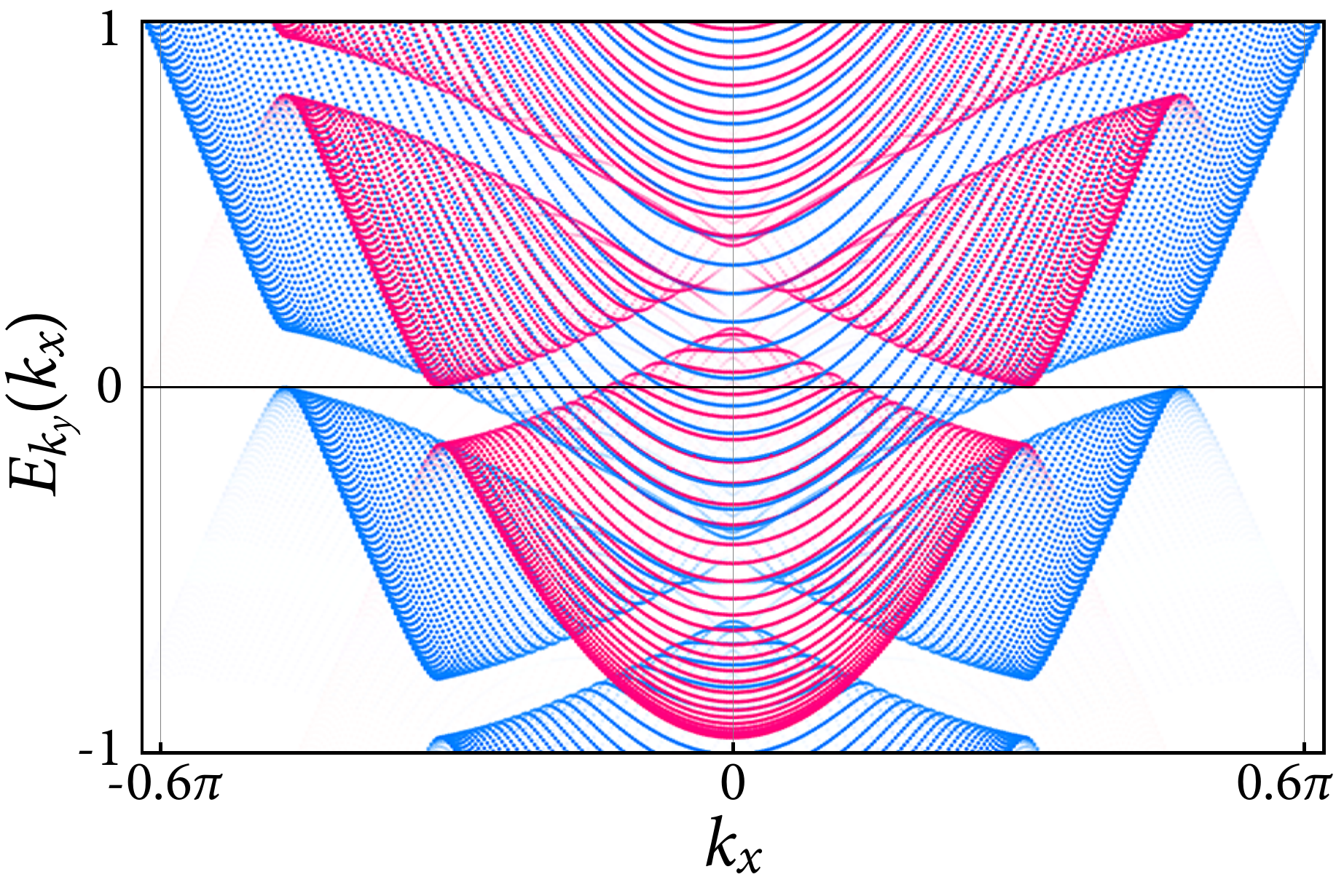}
\put(-5,63.5){(d)}
\end{overpic}
\vspace{-1.5mm}
\caption{(Color online) Gap structure of the superconducting solutions of Table~\ref{Tab1}. For each $k_x$, the eigenenergies $E_{k_x}(k_y)$ are plotted as a function of $k_y$ (except for the LO solution (d), where $E_{k_y}(k_x)$ is plotted). The eigenenergies of band 1 and band 2 are plotted in pink and in blue, respectively. The opacity of each point encodes the weight with which the corresponding state contributes to the density of states, given through the coherence factors.}
\label{Fig3}
\end{figure*}

In the presence of at least two different COMMs, the system is inhomogeneous with finite off-diagonal elements of the density matrix ~\cite{loder10}. In the presence of the two COMMs $\q_1$ and $\q_2$, it exhibits a charge-stripe order with wavevector $\q_1-\q_2$. Within the mean-field decoupling of the interaction (\ref{g2.1}), additional order parameters, representing charge order, could be defined. However, their amplitude remains small for the values of the COMMs relevant here, and therefore we subsequently neglect these order parameters.

\section{Results}%

\begin{figure}[htb]
\begin{overpic}
[height=0.465\columnwidth]{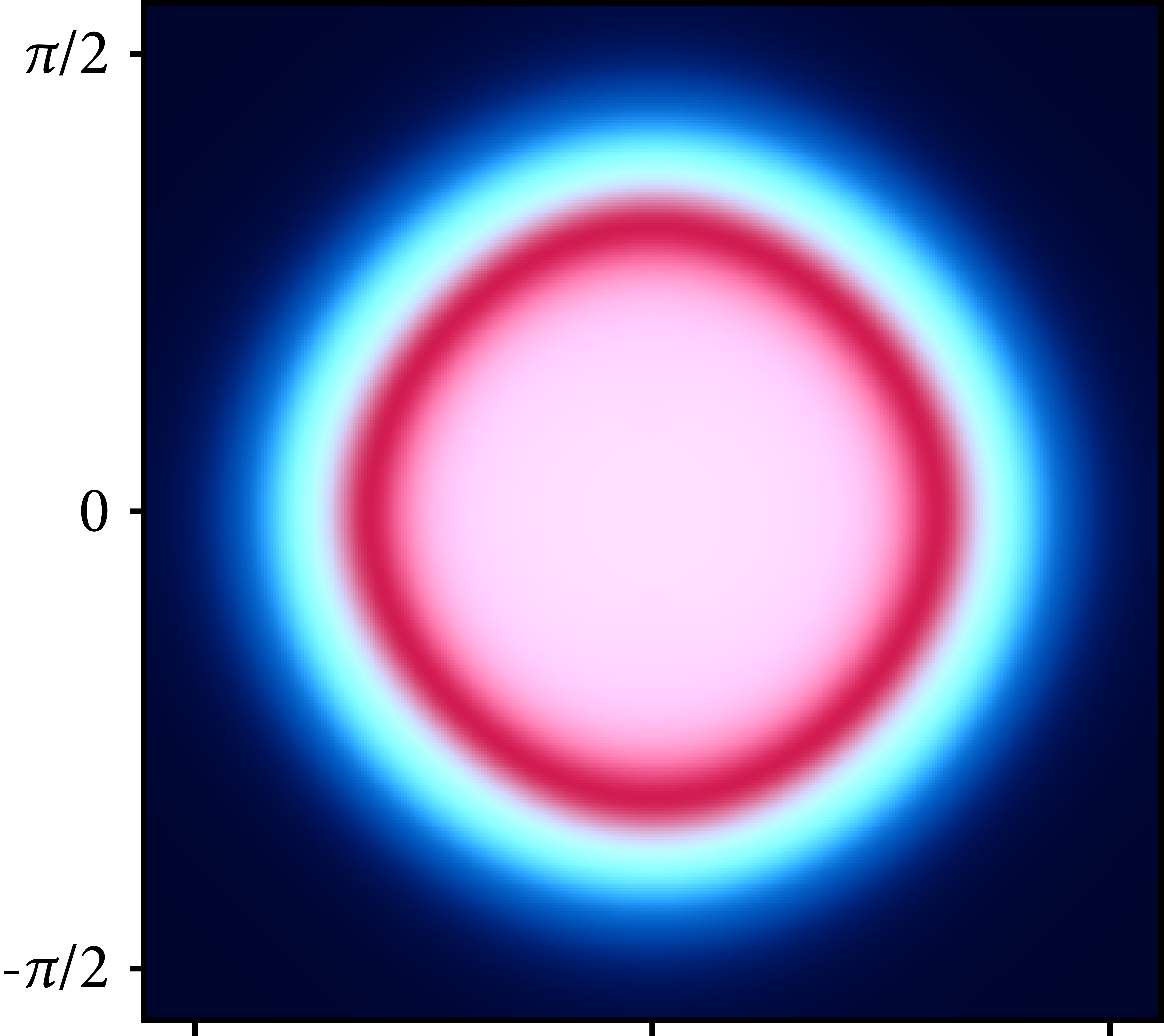}
\put(14.5,80){\color{white}\large$\bm{n(\k)}$}
\put(87.5,5){\color{white}\large\bf a1}
\end{overpic}\hspace{-5mm}
\begin{overpic}
[height=0.465\columnwidth]{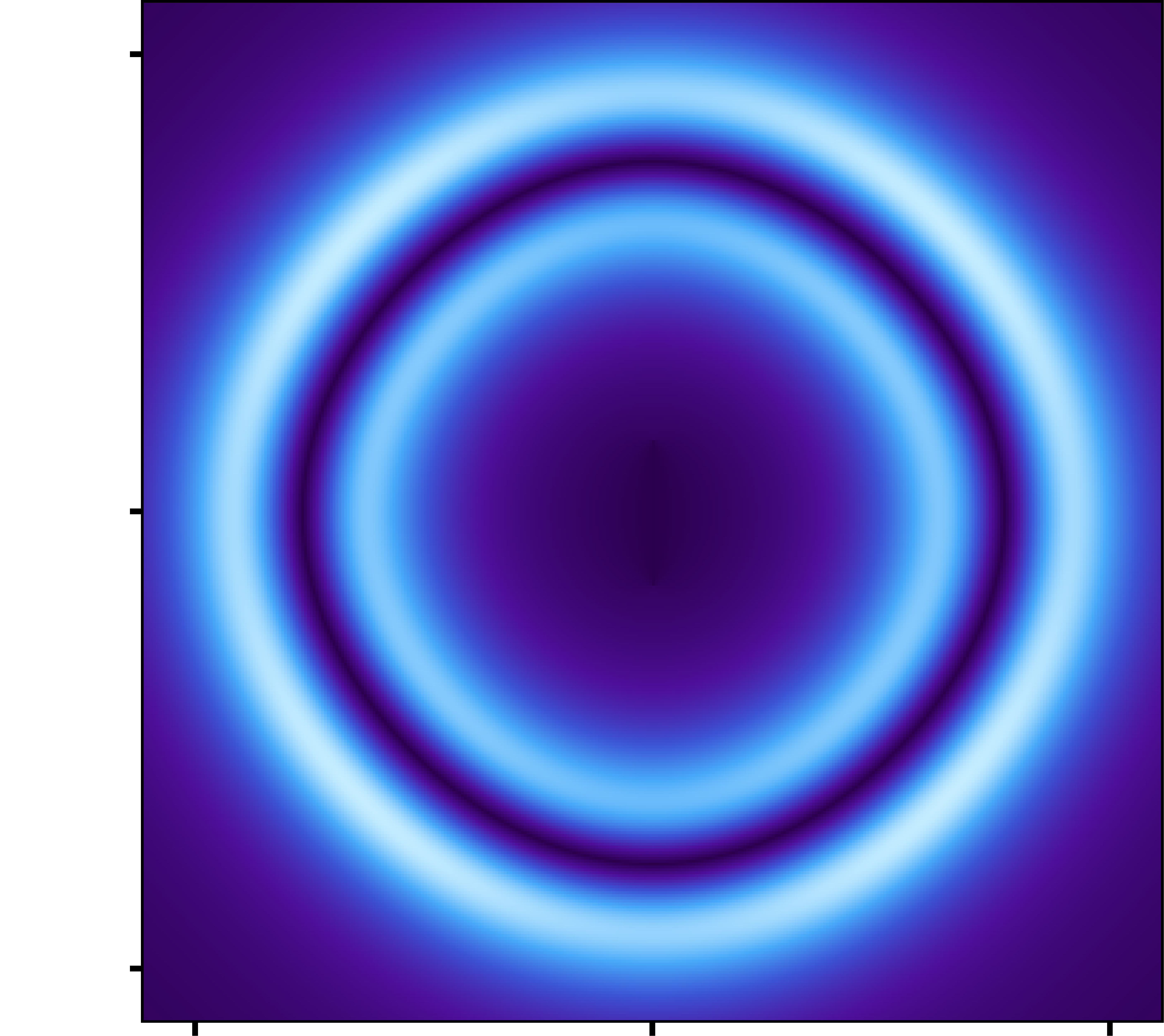}
\put(14.5,80){\color{white}\large$\bm{p(\k)}$}
\end{overpic}
\begin{overpic}
[height=0.465\columnwidth]{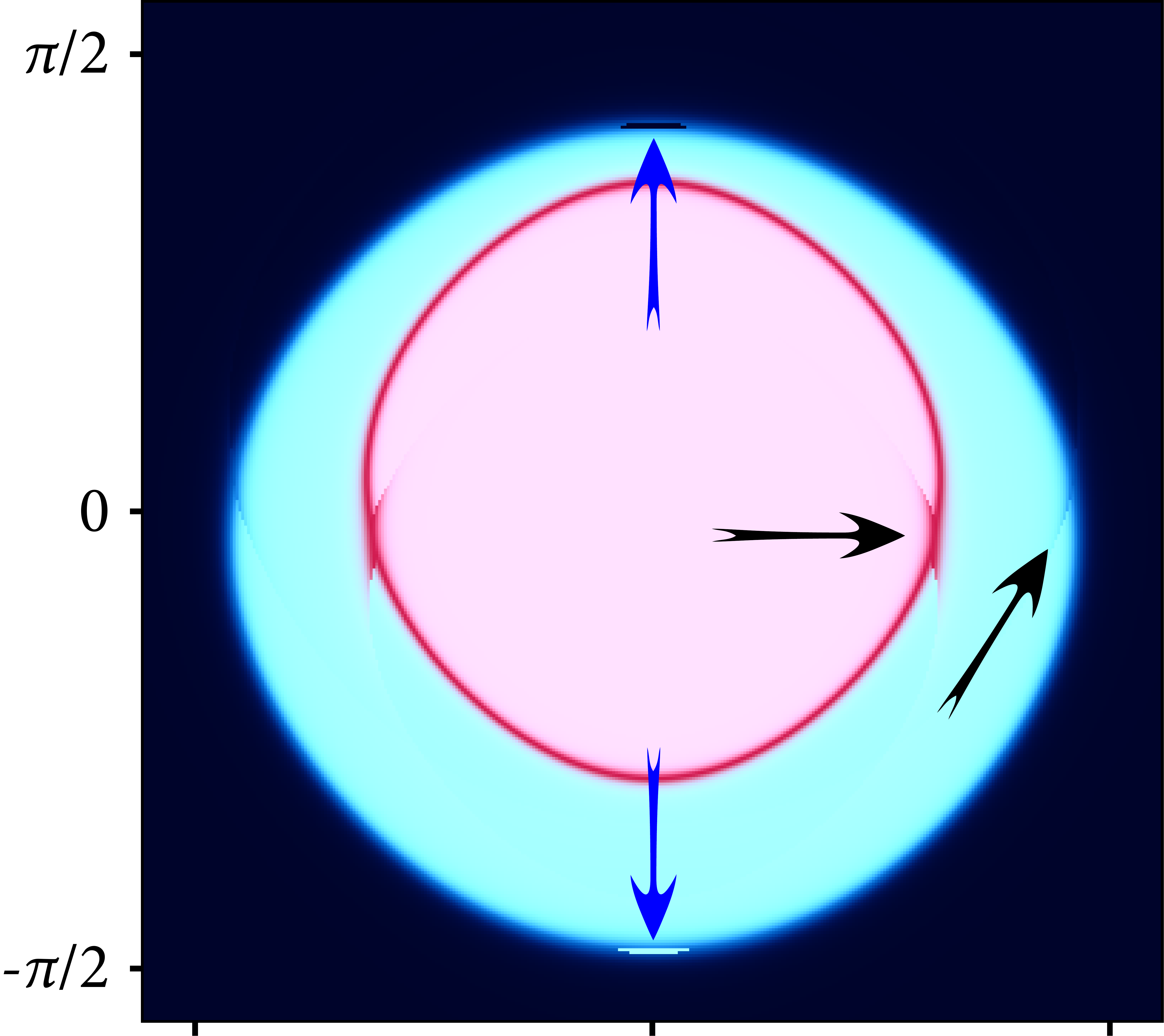}
\put(87,5){\color{white}\large\bf b1}
\end{overpic}\hspace{-5mm}
\begin{overpic}
[height=0.465\columnwidth]{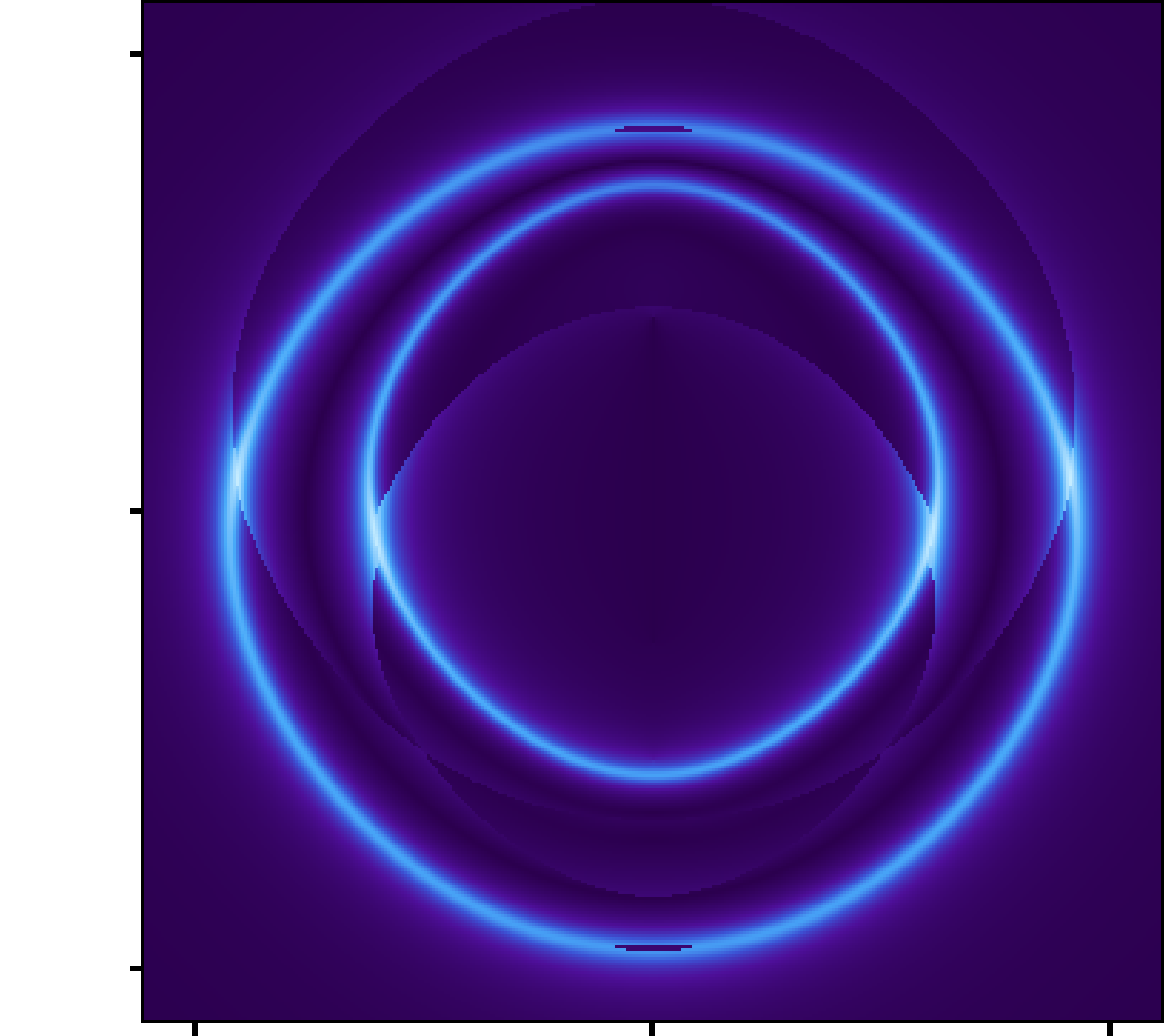}
\end{overpic}
\begin{overpic}
[height=0.465\columnwidth]{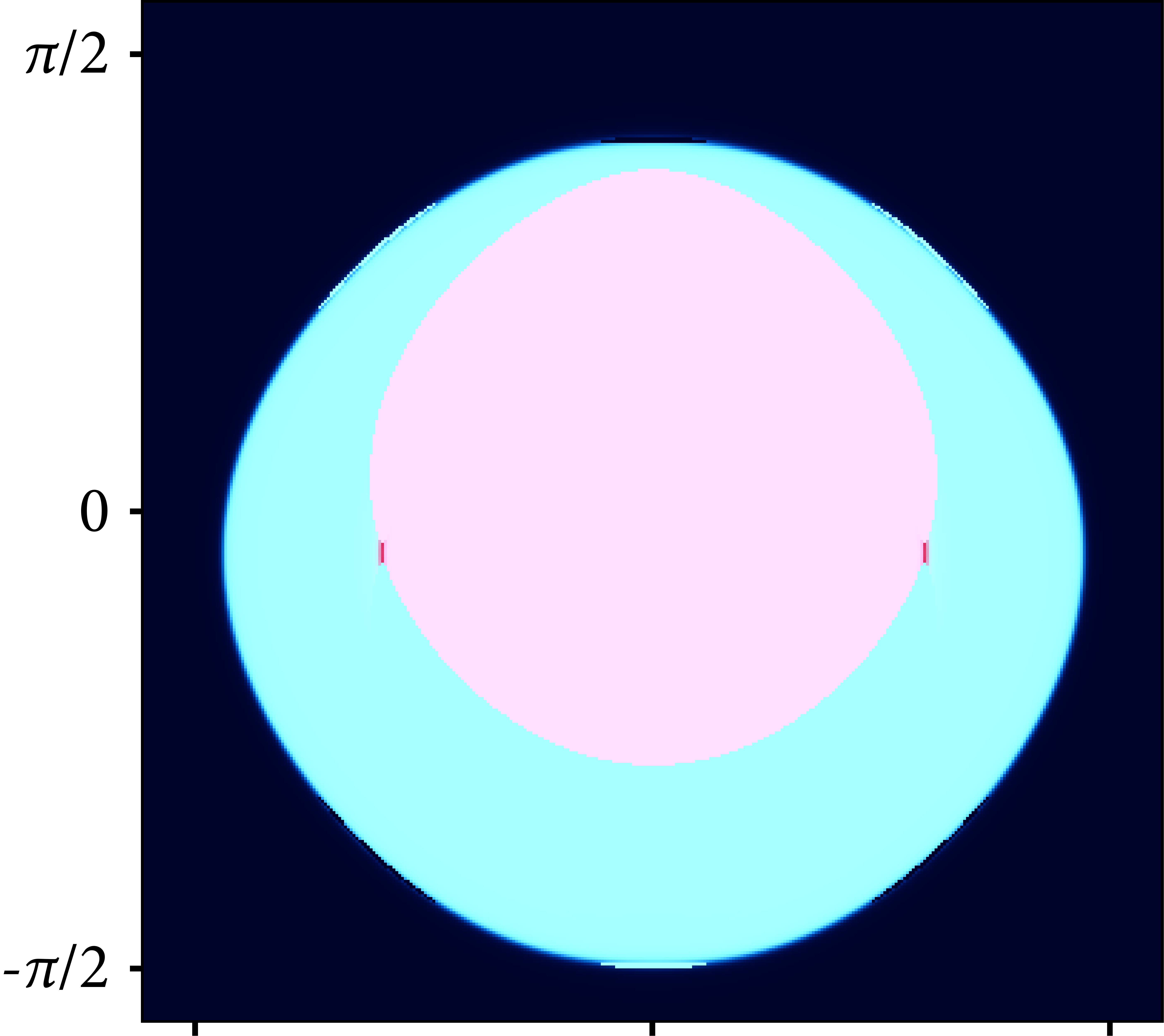}
\put(87,5){\color{white}\large\bf b2}
\end{overpic}\hspace{-5mm}
\begin{overpic}
[height=0.465\columnwidth]{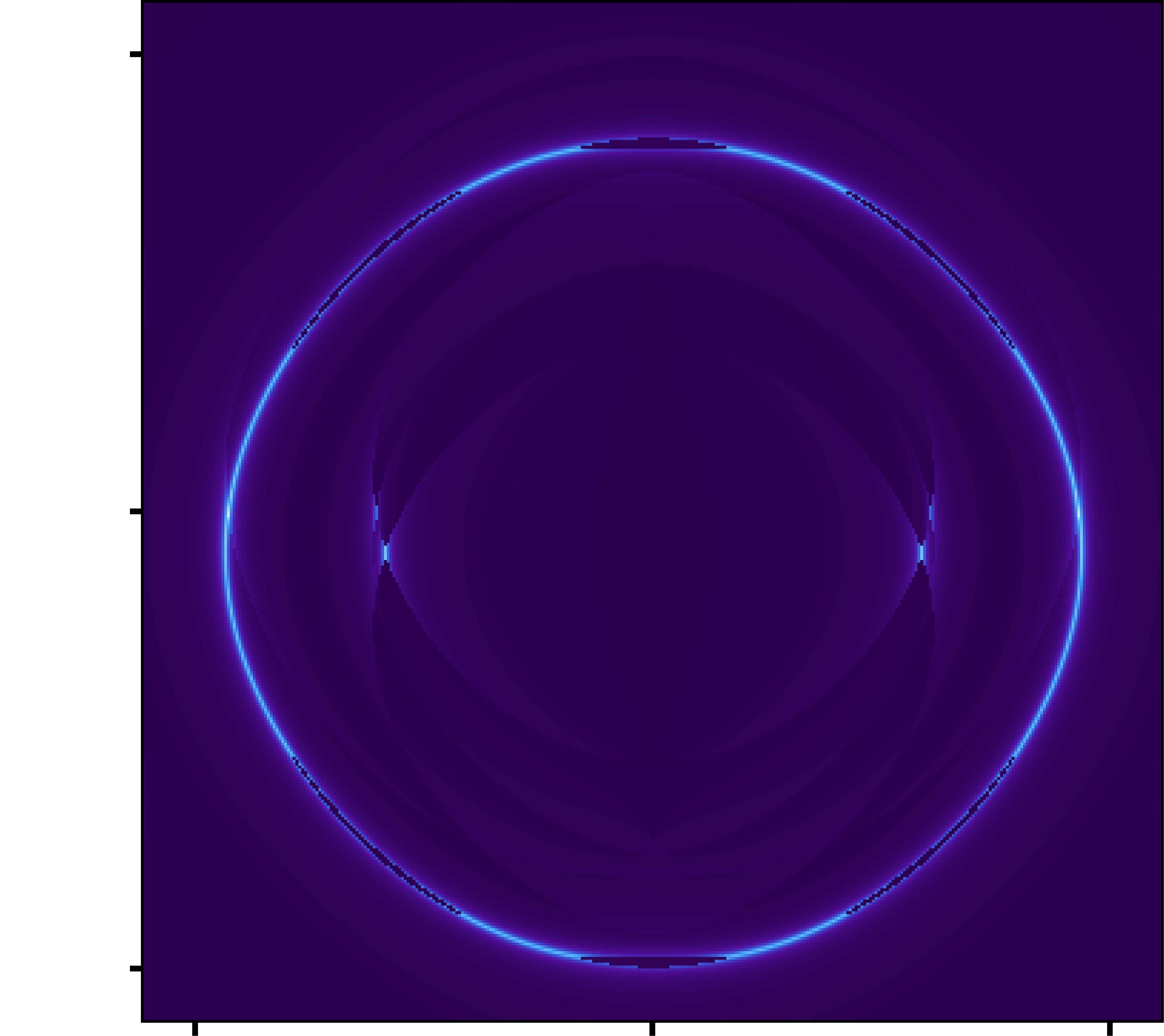}
\end{overpic}
\begin{overpic}
[height=0.465\columnwidth]{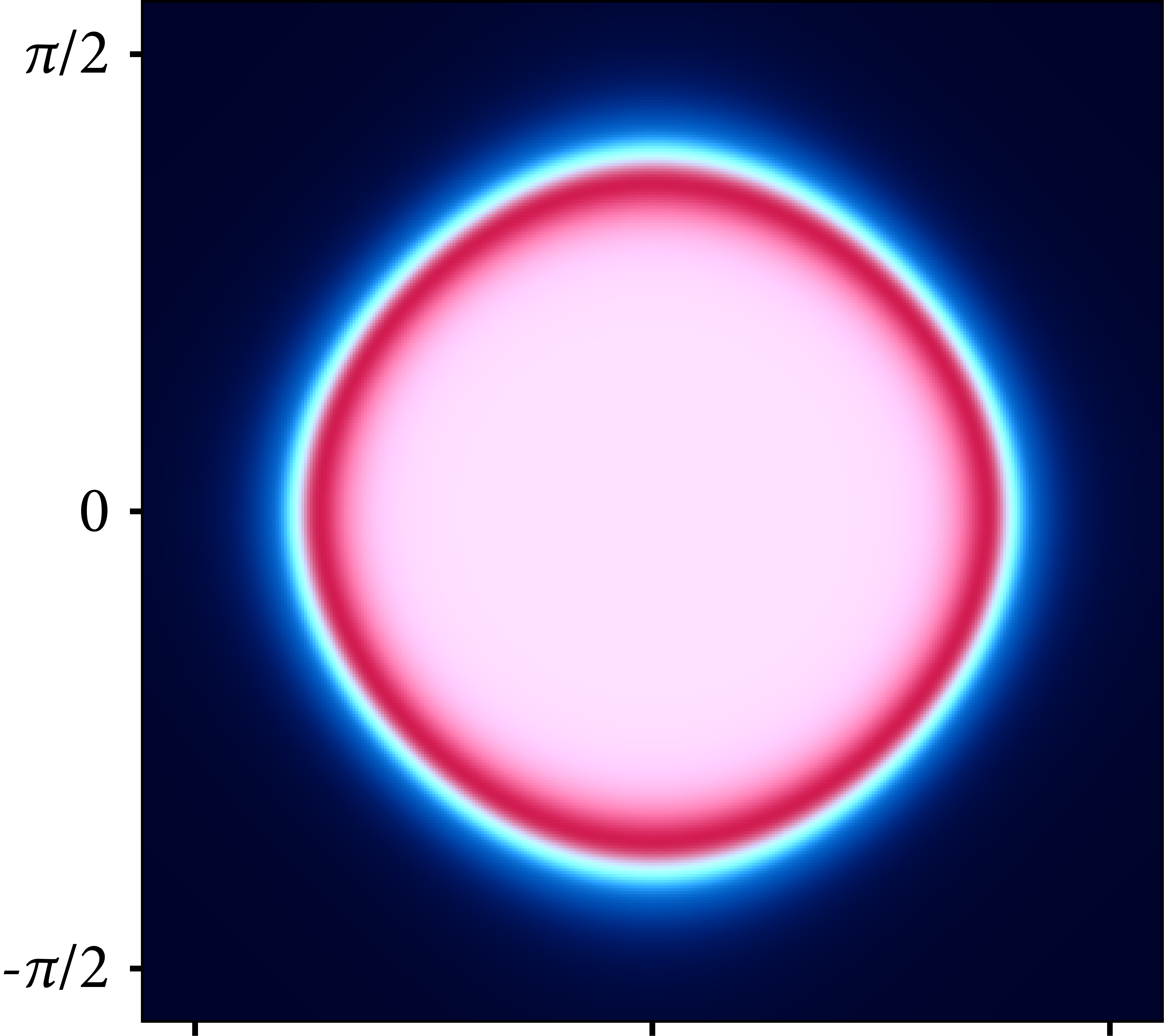}
\put(88,5){\color{white}\large\bf c}
\end{overpic}\hspace{-5mm}
\begin{overpic}
[height=0.465\columnwidth]{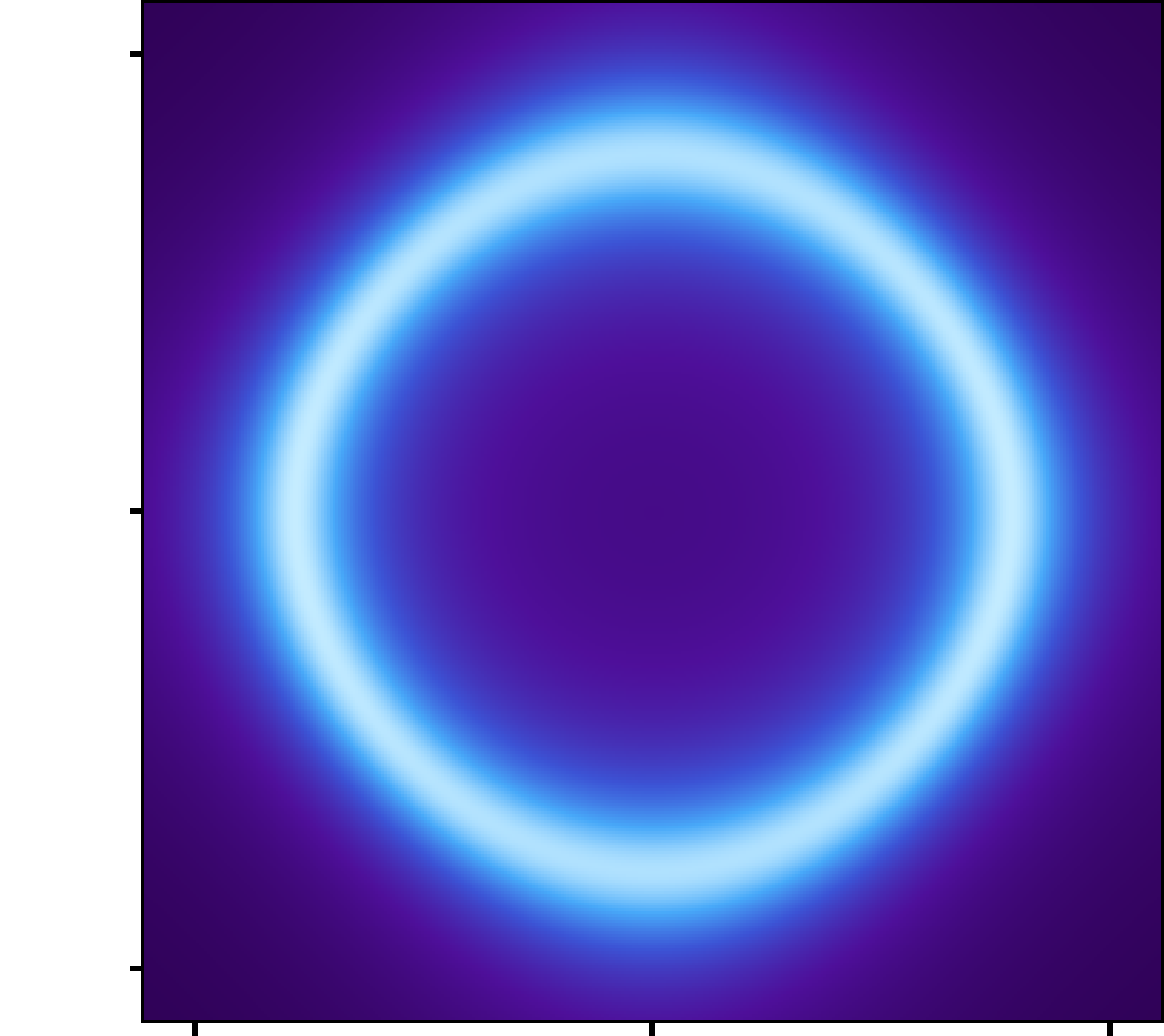}
\end{overpic}
\begin{overpic}
[height=0.502\columnwidth]{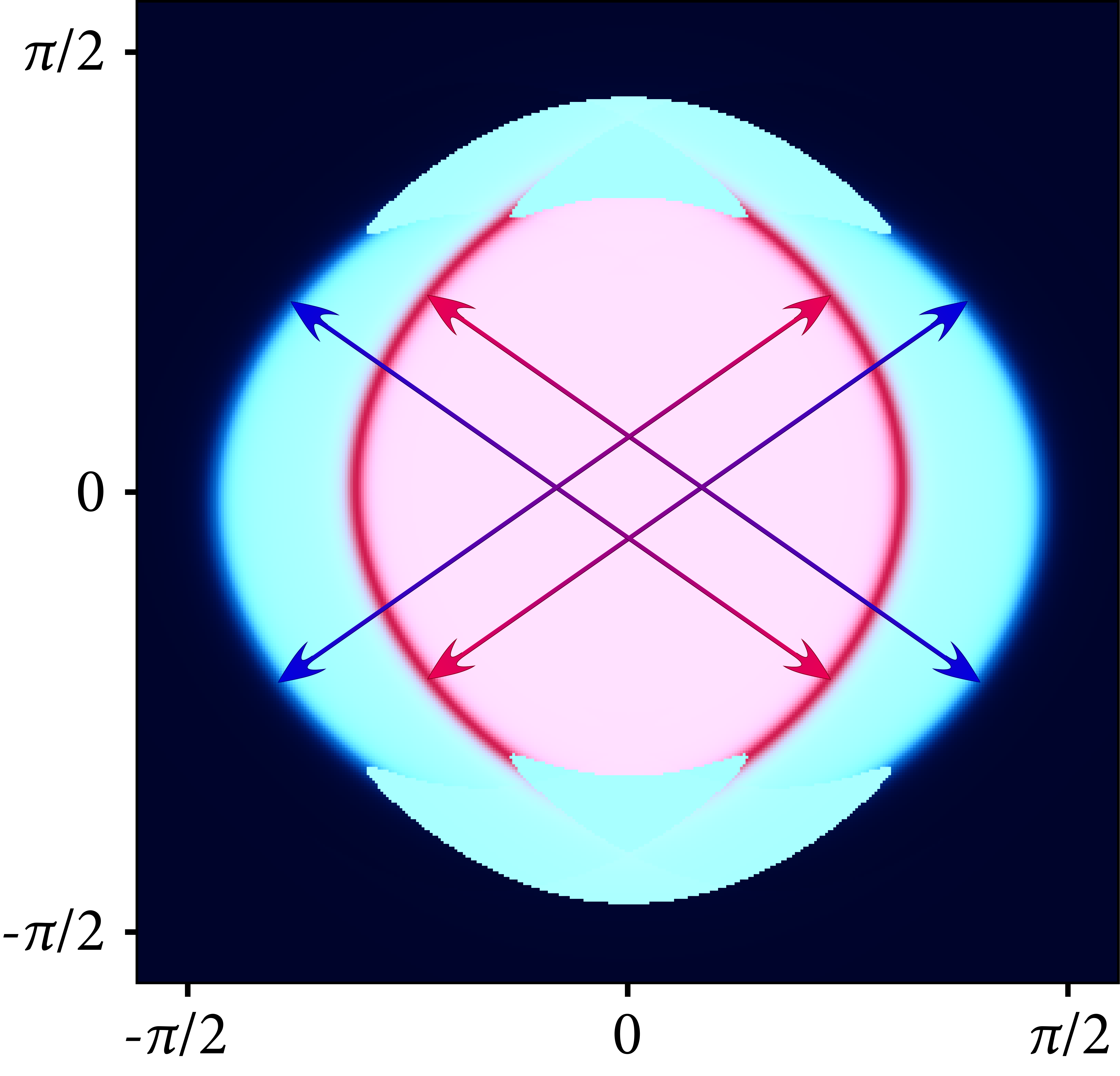}
\put(88,11.5){\color{white}\large\bf d}
\end{overpic}\hspace{-5mm}
\begin{overpic}
[height=0.502\columnwidth]{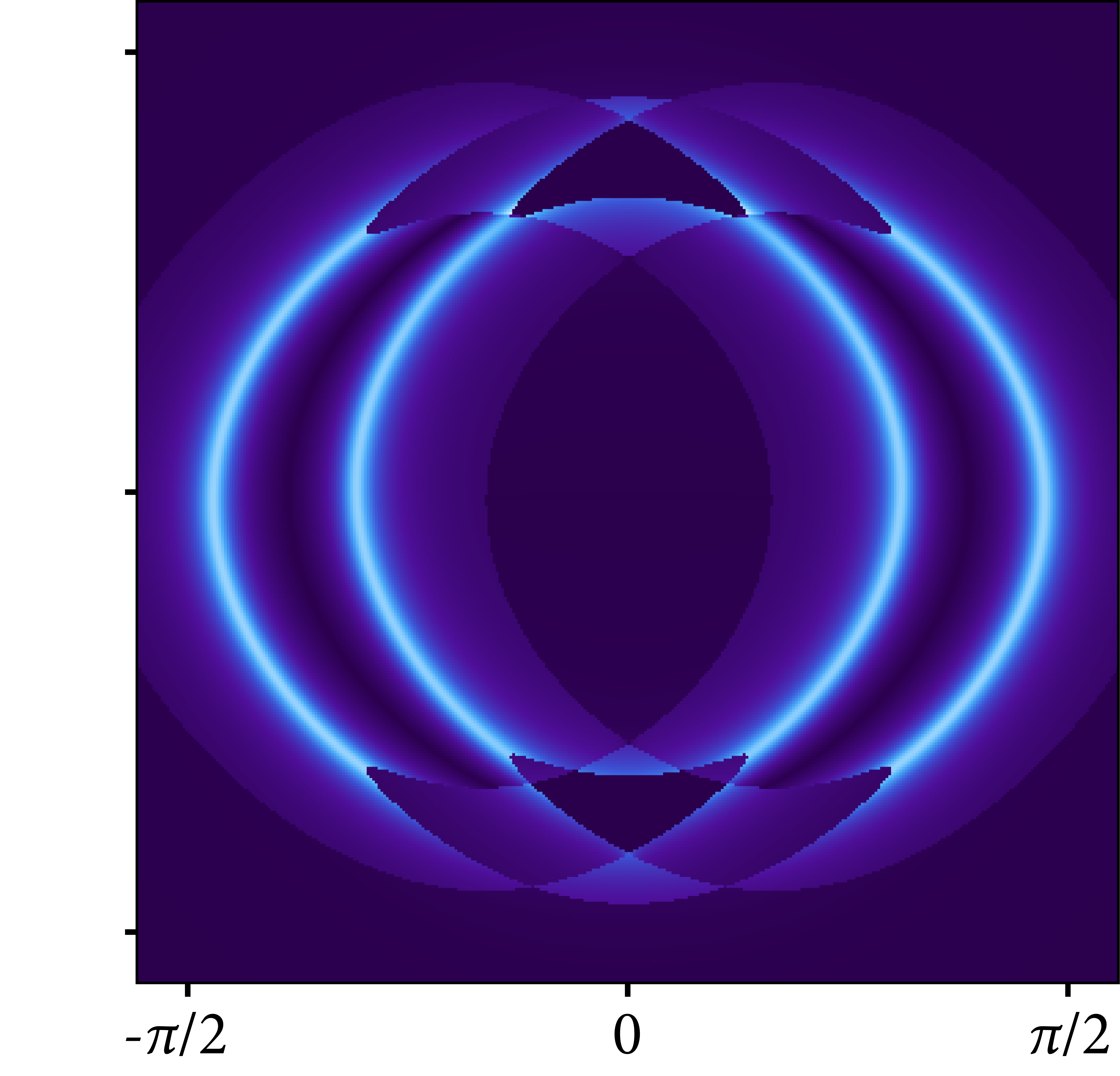}
\end{overpic}\vspace{2mm}\\
\makebox[3.6mm]{}
\begin{overpic}
[height=0.08\columnwidth]{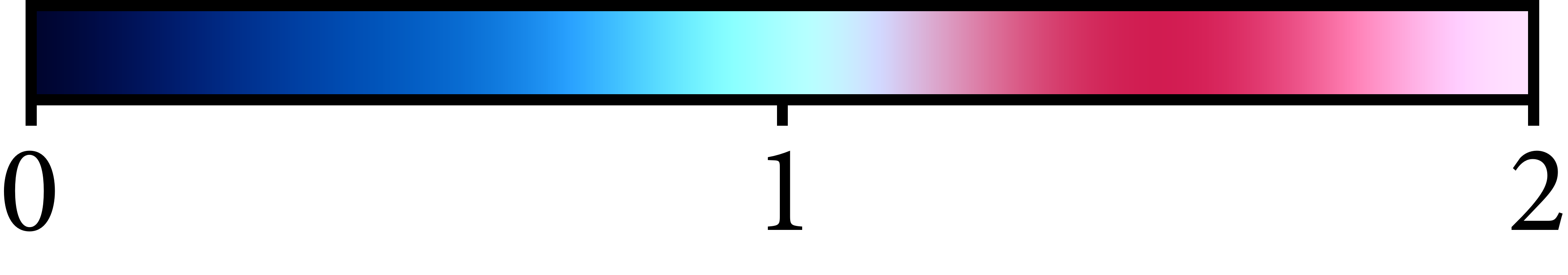}
\end{overpic}
\makebox[0.0mm]{}
\begin{overpic}
[height=0.08\columnwidth]{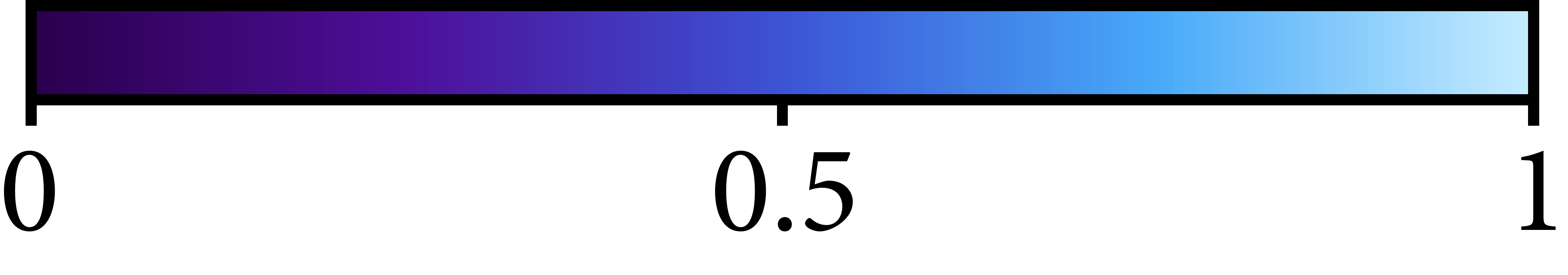}
\end{overpic}
\vspace{-1.5mm}
\caption{ $n(\k)$ (left panel) and $p(\k)$ (right panel) for the solutions a1) -- d) of Tab.~\ref{Tab1}. For the arrows in b1) and d) see main text.}
\label{Fig4}
\end{figure}

We choose the parameters $n=0.24$ and $V=1.7\,t$, for which stable solutions for all the different SC phases exist. Figure~\ref{Fig2} shows the phase diagram in the $\alpha$--$H_x$ plane. As expected, intra-band pairing dominates throughout the half-plane $\alpha>H_x$, although superconductivity is steadily suppressed by an increasing magnetic field. If $H_x>0$, the normal-state Fermi surfaces shift as shown in Fig.~\ref{Fig1} and two eigenstates on opposite points of one Fermi surface differ in energy by $\Delta E(\k)\approx|\xi_\alpha(\k_{\rm F})-\xi_\alpha(\k_{\rm F}-{\bf K}_\alpha)|$. Nevertheless, if $\Delta E(\k)<\Delta^\text{odd}(\q=\bm0)$ for all $\k$, pairing with zero COMM is favored (phase a). In lowest order in $H_x$, one finds that the maximal value of $\Delta E(\k)$ (for $k_x=0$) is $2\alpha H_x/\sqrt{\alpha^2+4t|\mu|}$. If this value exceeds $\Delta^\text{odd}(\bm0)$, only solutions with finite COMM pairing exist (phase b). This phase is similar to the ``helical phase'' of Ref.~[\onlinecite{kaur05}], although in our calculations two COMMs $\q_1\approx2{\bf K}_1$ and $\q_2\approx2{\bf K}_2$ coexist in the groundstate of this phase, optimizing pairing in band 1 and band 2, respectively (solution b1 in Tab.~\ref{Tab1}). Upon approaching the boundary of phase b, pairing on the smaller Fermi surface (of band 1) is suppressed (solution b2) and a one-band SC state remains.~\footnote{It is not possible to decide from our numerical calculations, whether a small pair density in band 1 survives all the way to the phase boundary, or whether an actual crossover to a one-band SC phase occurs. Compare the finite values of $\Delta_{11,22}(\q_1)$ and the distinct value of $\q_1$ in Tab.~\ref{Tab1}.} The finite pair density in band 2 induces an energy gap also in band 1. This ``interior gap'' phenomenon was predicted also in two-band systems with strongly differing effective electron masses~\cite{liu03}; in our model however, the induced energy gap is momentum dependent (see discussion of Fig.~\ref{Fig3} below).

Inter-band pairing is possible below $\alpha=H_x$. The phase transition at $\alpha=H_x$ is first order and typically ends at a critical point, above which intra- and inter-band pairing are separated by a non-superconducting metallic phase. The position of this critical point changes with system parameters such as $V$, $t'$ and the electron density, and occurs at $\alpha=H_x=0.082\,t$ in Fig.~\ref{Fig2}.
At the critical magnetic field $H_{\rm LO}(\alpha)$ (with $H_{\rm LO}(0)=\Delta^\text{even}(\q=\bm0)$), a first-order transition into the LO phase takes place, which is stable within a certain field range above $H_{\rm LO}$ and well separated from intra-band pairing. The extent of the SC phases on the $H_x$ axis scales with the interaction strength $V$. The qualitative structure of the phase diagram remains however unchanged upon changing $V$.

In the following we discuss the characteristic properties of the two types of intra- and the inter-band pairings on the basis of four typical parameter sets, marked with white stars in Fig.~\ref{Fig2} and summarized in Table~\ref{Tab1}. As already mentioned above, superconducting states with at least two coexisting COMMs are generally inhomogeneous, i.e., states with COMMs $\q_1$ and $\q_2$ form a ``pair-density wave'' with wave vector $(\q_1-\q_2)/2$ and a concomitant charge-density wave with a wave vector $\q_1-\q_2$~\cite{berg09,loder10}.
The properties of the different phases are essentially determined by the structure of the momentum dependent energy gap resulting from the OPs $\Delta^\text{even(odd)}(\k,\q)$. This is analyzed in Fig.~\ref{Fig3}, showing the eigenenergies $E_{k_x}(k_y)$ for all $k_x$ as a function of $k_y$ (except Fig.~\ref{Fig3}~(d) showing $E_{k_y}(k_x)$). Figure~\ref{Fig3}~(a1) shows the zero COMM intra-band pairing state with an almost uniform energy gap $\Delta^\text{odd}(\k,\bm0)$ in both bands. Because the normal-state Fermi surfaces are shifted in $k_y$-direction, the ``gaps'' are tilted as a function of $k_y$. As a consequence, the coherence peak in the density of states (DOS) splits into a double-peak structure.

Upon increasing $\alpha$ or $H_x$, the displacement of the Fermi spheres and therefore the tilting of the energy gaps increases, reaching a point where the upper and lower bands touch [Fig.~\ref{Fig3}~(a2)]. Beyond this limit, zero COMM superconductivity breaks down and is replaced in a first-order transition by a finite COMM state [Fig.~\ref{Fig3}~(b1)]. In this state, pairing with finite COMM $\q_{1,2}\approx2{\bf K}_{1,2}$ compensates the shift of band 1 or of band 2, respectively, and generates an energy gap covering the full Fermi surface of band 1 or of band 2. On the other hand, these pairings also induce an interior gap in the other band, which results in an overall energy cost, because the interior gap shifts more eigenstates to higher energies than to lower energies. Therefore, upon reaching the outer phase boundary of the SC phase b, pairing on the smaller Fermi surface (band 1) is lost, leaving only the gap in band 2 and the induced interior gap in band 1 [Fig.~\ref{Fig3}~(b2)].

The gap structure of the zero COMM inter-band pairing state is shown in Fig.~\ref{Fig3}~(c). Clearly visible is a gap shifted to higher energy in band 1 and a gap shifted to lower energy in band 2. At the critical magnetic field $H_{\rm c}$, the gap edges in band 1 and band 2 touch, thus for magnetic fields above $H_{\rm c}$, zero COMM pairing is no longer possible. The LO state above $H_{\rm c}$ [Fig.~\ref{Fig3}~(d)] has a gap structure similar to the finite COMM intra-band pairing state. However, because of the required inter-band pairing, there is no COMM generating an energy gap which covers the whole Fermi surface of one band. Instead, the COMMs $\pm\q$ of the LO state are equal to the momentum difference of the Fermi surfaces of band 1 and band 2 (c.f.~Fig.~\ref{Fig4}~(d), where paired states are indicated by the two-colored arrows). Each band therefore contains two tilted energy gaps for $+\q$ and $-\q$, which overlap the Fermi energy only partially. The direction of the COMMs of the LO state is not a priori clear. While the states with $\q$ in $x$- and in $y$-direction are equivalent for $\alpha=0$, in the calculations with $\alpha=0.05\,t$ (see Table~\ref{Tab1}), COMMs parallel to $\bf H$ are favored.

A complementary characterization of the distinct SC phases is given by the momentum distribution 
\begin{align}
n(\k)=\langle c^\dag_{\k\ua}c_{\k\ua}\rangle+\langle c^\dag_{\k\da}c_{\k\da}\rangle
\end{align}
(left panels of Fig.~\ref{Fig4}) and the pair density
\begin{align}
p^2(\k)=\sum_\q|\langle c_{-\k+\q\da}c_{\k\ua}\rangle|^2
\end{align}
(right panels of Fig.~\ref{Fig4}). Fermi surfaces are represented by lines of discontinuities in $n(\k)$, which becomes continuous wherever the Fermi energy falls into an energy gap. The pair density $p(\k)$ is typically largest where the derivative of $n(\k)$ is largest (these are the dark red and dark blue regions in $n(\k)$), but it is zero where $n(\k)$ changes discontinuously. Pairing occurs only where an energy gap crosses the Fermi energy, but no pairing occurs at interior gaps, since these do not influence $n(\k)$. Interior gaps are induced by pairing on the Fermi surface of the other band and do not contribute to the condensation energy. In the intra-band pairing states (a1, a2, b1, and b2), pairing occurs between two blue quasi-particle states and between two red quasi-particle states, whereas in the inter-band pairing states [c and d], pairing occurs between a red and a blue quasi-particle state (indicated by the two-colored arrows in Fig.~\ref{Fig4}~d).

\begin{figure}[t]
\centering
\vspace{3mm}
\begin{overpic}
[width=0.95\columnwidth]{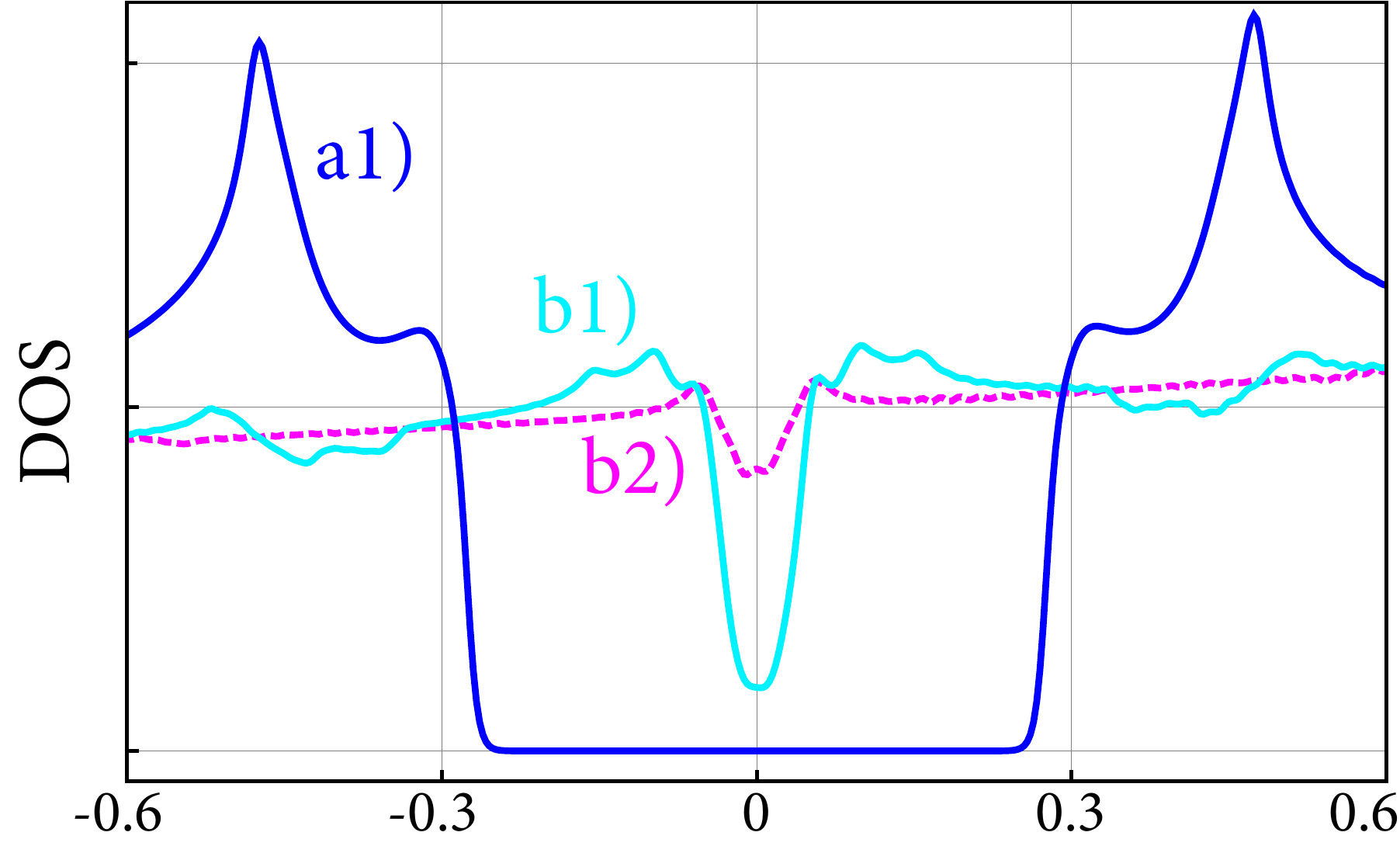}
\end{overpic}\vspace{2mm}
\begin{overpic}
[width=0.95\columnwidth]{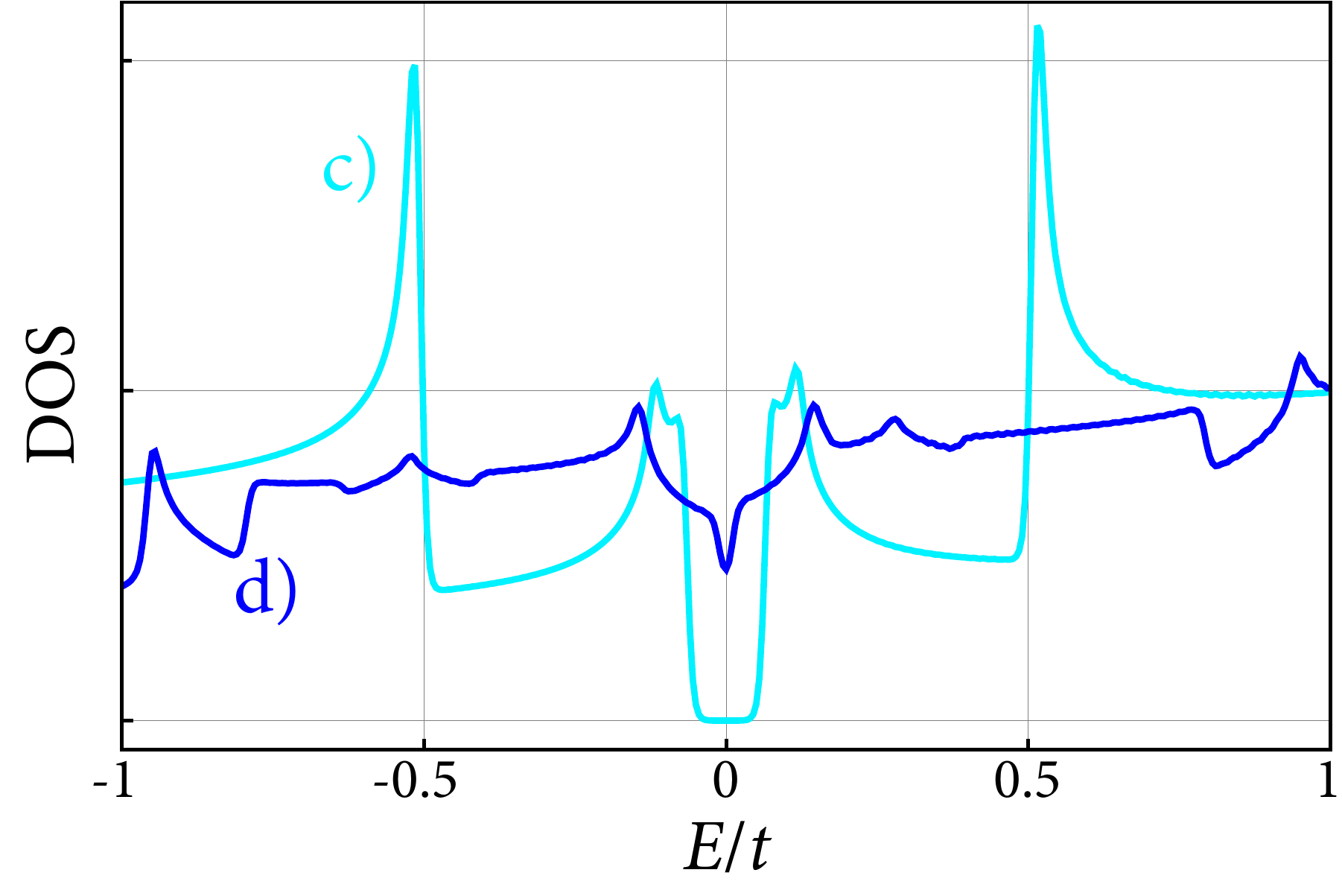}
\end{overpic}
\vspace{-1.5mm}
\caption{(Color online) Density of states as a function of the energy relative to the Fermi level for dominating SOC (upper panel) and dominating magnetic field (lower panel). Upper panel: intra-band solutions a1) (dark blue), b1) (light blue), and b2) (dashed purple) of Table~\ref{Tab1}. Lower panel: inter-band solution c) (light blue) and LO solution d) (dark blue) of Table~\ref{Tab1}.}
\label{Fig5}
\end{figure}

The low-energy physics is characterized essentially by the OP $\Delta^\text{even/odd}(\q)$.
Most eminent is the fact that $n(\k)$ and $p(\k)$ are almost $x$--$y$ symmetric in both zero COMM pairing phases (a1 and a2), although $x$--$y$ symmetry is clearly broken in the normal state. On the other hand, the finite COMM SC states (b1, b2, and d) break $x$--$y$ symmetry. This change of symmetry upon increasing the magnetic field might be observable experimentally, as we will argue below.

We have thus identified four distinct SC phases, the different symmetry properties of which are most evident in their corresponding pair densities $p(\k)$ (right panels of Fig.~\ref{Fig4}). Characteristic for the finite COMM phases is also a fascinating fine structure in both $n(\k)$ and $p(\k)$, originating from partially reconstructed Fermi surfaces (marked by black arrows in Fig.~\ref{Fig4} b1) and Fermi pockets (blue arrows in Fig.~\ref{Fig4} b1), which are typically present in an inhomogeneous SC system with two different COMMs~\cite{baruch08,berg09,loder10}. These structures are most likely not observable experimentally. A characteristic quantity in which they may leave fingerprints is the DOS (Fig.~\ref{Fig5}). For the chosen band filling ($n=0.24$), the normal-state DOS is featureless near the Fermi energy. Zero COMM pairing typically generates a full energy gap, although a double-peak structure in the coherence peaks appears in a finite magnetic field. This double peak originates from the tilting of the dispersion. At the critical magnetic field $H_{\rm LO}$, the inner coherence peaks touch and the gap closes. Characteristic for finite COMM pairing is a finite DOS at the Fermi energy and additional structure away from the Fermi energy which originates from the interior gaps. In the LO state, parts of the Fermi surface around $k_x=0$  do not lie within the energy gap and therefore survive into the SC state. The origin of the in-gap states in the finite COMM intra-band pairing phase is more complex; they originate from the partially reconstructed Fermi surfaces. These in-gap states are visible in Fig.~\ref{Fig3}~(b1) close to $k_y=0$ and are responsible for the lines of discontinuities crossing the dark red and dark blue ``circles''  in Fig.~\ref{Fig4}~(b1) (marked by the black arrows)~\footnote{This effect is also present in the LO phase, although in this phase more in-gap states arise from the tilted energy gaps.}.

\section{Discussion}
A prominent example, in which the combination of superconductivity with SOC and magnetism is relevant, is the superconducting LAO-STO interface. We therefore estimate which SC phases may occur in this system.
The parameters characterizing the LAO-STO samples vary considerably in different experiments. Here we use the values given by Caviglia {\it et al}. for the density of mobile electrons at the interface and the Rashba SOC~\cite{caviglia08,caviglia10}. The electron density for zero gate voltage is $4.5\times10^{13}\,\text{cm}^{-2}$, which corresponds to $n\approx0.07$ mobile electrons per unit cell. For a quadratic dispersion relation one obtains $E_{\rm F}\approx35\,$meV and the energy scale $t\approx80\,$meV. 
The measured values of the Rashba SOC strength $\alpha_\text{exp}=a\alpha$ (where $a$ is the lattice constant of LAO) vary with the gate voltage from 1 to 5$\times 10^{-12}\,$eVm~\cite{caviglia10}. In our model, these values correspond to $\alpha\approx0.03$ -- $0.15\,t$.

The intrinsic magnetism at the interface is strongly inhomogeneous, thus no definite position on the $H_x$ axis can be given. The average in-plane magnetic moment varies between $0.01\mu_{\rm B}$~\cite{bert11} and $0.3\mu_{\rm B}$ per unit cell~\cite{li11}. This corresponds to $\sim15$ -- 500\,G, or to $H_x\approx0.1$ -- $3\times10^{-5}\,t$ in our model. 
For all of these values, the system is therefore deep in the zero COMM intra-band phase. For $\alpha\gg H_x$, the phase boundary to the finite COMM pairing phase occurs at $H_x\approx\Delta\sqrt{\alpha^2+4t|\mu|}/2\alpha\approx0.05$$-$$0.25H_x$, where $\Delta$ is the SC energy gap. From measurements of the superfluid density~\cite{bert12}, an estimate $\Delta\approx 40\,$\textmu eV can be deduced. The in-plane magnetic field required for the finite COMM phase (phase b) is therefore around 1--5\,T, which is far above the measured internal magnetic field, but can easily be reached in the laboratory. The LAO-STO interface is therefore a likely candidate for the observation of a finite COMM superconducting state.

\begin{figure}[t]
\centering
\vspace{3mm}
\begin{overpic}
[width=\columnwidth]{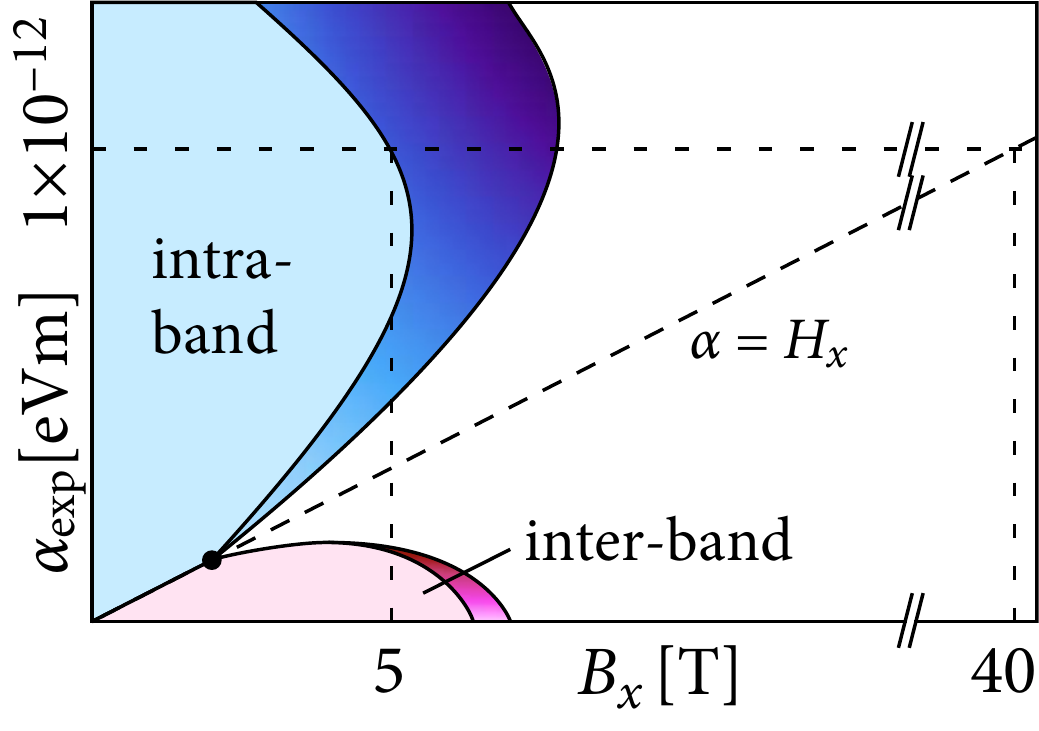}
\end{overpic}
\vspace{-3mm}
\caption{(Color online) Sketch of the phase diagram for the experimentally obtained parameters of the LAO-STO interface system (see main text). Here, $B_x=H_x/\mu_{\rm B}$ and $\alpha_\text{exp}=a\alpha$, where $a$ is the lattice constant of STO. For the smallest reported value of $\alpha_\text{exp}$ ($10^{-12}$\,eVm), we estimate the transition into the finite COMM intra-band pairing phase to occur at $\approx5$\,T, while $\alpha=H_x$ corresponds to $\approx40$\,T. A reentrance into the inter-band pairing phase upon increasing $H_x$ is therefore not possible.}
\label{Fig5}
\end{figure}

By applying an even larger external magnetic field one possibly might wonder whether this system can be moved into the inter-band phase. The magnetic field $H_x=\alpha$ corresponds to $\sim40\,$T for $\alpha=0.03\,t$. According to our discussion above, this is the minimal field above which this inter-band pairing is possible. However, this field is far larger than the critical field $H_{\rm LO}=\Delta^\text{even}(H_x=0)\approx40\,$\textmu eV. For the LAO-STO interface, the energy gap is therefore $\sim10^2$ -- $10^3$ times smaller than the spin-orbit band splitting reported in Ref.~[\onlinecite{caviglia10}] (and the order parameters obtained in Table~\ref{Tab1}) and a reentrance into the inter-band superconducting state is not possible for this system. The phase diagram obtained from our model for the experimentally determined parameters of the LAO-STO interface system is thus quantitatively different from the one shown in Fig.~\ref{Fig2}. In particular, superconductivity extends only little in the direction of the magnetic field and the normal state is reached for magnetic fields $H_x\ll\alpha$. This situation is schematically shown in Fig.~\ref{Fig5}.

Theoretical estimates for $\alpha$ are difficult. A simple electrostatic model for the interface, using Dirac's result $\alpha=|E|e\hbar^2/4m^{*2}c^2$ yields a value for the Rashba SOC strength that is many orders of magnitude smaller than the values reported from the experiments. In this scenario, inter-band pairing is expected even in moderate external magnetic fields. However, for electron densities above $\sim1.5$--$1.9\times10^{13}\,\text{cm}^{-2}$, a multi-band behavior has been inferred from Hall effect measurements~\cite{joshua:11}, for which the effective value of $\alpha$ can be strongly enhanced~\cite{joshua:11,zhong:13}. By tuning the electron density at the interface through a gate voltage~\cite{thiel06}, both the one-band and the multi-band regime may be accessible~\cite{joshua:11,joshua:12}. Therefore, superconductivity with weak and strong SOC may both be realized at the LAO-STO interface, depending on the gate-voltage tuned electron density.

Because of the uncertainties in the parameter values and the sensitivity of our model to the strength of the magnetic field, it is difficult to predict details in the structure of the DOS. For the reported internal magnetic fields, a double-peak structure is unlikely to be resolvable experimentally, and was indeed not seen in Ref.~[\onlinecite{bert12}].

In some heavy-fermion superconductors like CePt$_3$Si or CeIrSi$_3$ inversion-symmetry is intrinsically broken~\cite{pfleiderer:09}. The physics discussed above may emerge in thin films of these materials in an external magnetic field, but is altered by vortices in the presence of a perpendicular field component~\cite{kaur05}. In the case of CePt$_3$Si, the SOC band splitting is of the order of 100\,meV~\cite{samokhin:2004} and exceeds the size of the SC energy gap by several orders of magnitude. This system would realize therefore the intra-band pairing phase and again, no reentrance into the inter-band pairing phase is possible. A correct description of the heavy-fermion superconductors is however more complicated, since their pairing interaction is non-local. This allows also for equal-spin triplet pairing~\cite{frigeri:04} and will therefore lead to a far richer phase diagram which has yet to be explored.

\begin{acknowledgements}
The authors gratefully acknowledge discussions with Julie Bert, Harold Hwang, Jochen Mannhart, Kathryn Moler, and Christoph Richter. This work was supported by the DFG through TRR 80.
\end{acknowledgements}

\vspace{5mm}

\end{document}